\documentclass[review]{elsarticle}

\usepackage{lineno,hyperref}
\modulolinenumbers[5]
\usepackage{comment}

\usepackage{graphicx}
\graphicspath{{./Images/}}

\newcommand{\fg}[1]{\mbox{\pmb{$#1$}}}
\newcommand{\bey}{\begin{eqnarray}}
\newcommand{\eey}{\end{eqnarray}}

\newcommand{\fvep}{\fg \varepsilon}

\newcommand{\bec}{\begin{center}}
\newcommand{\eec}{\end{center}}

\newcommand{\drop}[1]{}

\usepackage{amsbsy}
\usepackage{amsmath}
\usepackage{amsfonts}
\usepackage{amssymb}
\usepackage{textcomp}
\usepackage{cleveref}
\usepackage[margin=2.5cm]{geometry}
\usepackage{xcolor}
\usepackage{rotating}
\usepackage{float}
\usepackage{xr}
\usepackage[labelfont=bf]{caption}

\Crefname{figure}{\text{Fig.}}{\text{Figs.}}
\Crefname{equation}{\text{Eq.}}{\text{Eqs.}}
\biboptions{numbers,sort&compress}

\begin{document}
	
	\begin{frontmatter}
		
		
		\title{ Grain growth phenomenon during pressure-induced phase transformations at room temperature}
		

		
		\author{Valery I. Levitas\corref{cor1}\fnref{label1,label2,label3}}
		\ead{vlevitas@iastate.edu}
		\fntext[label1]{Department of Aerospace Engineering, Iowa State University, Ames, Iowa 50011, USA}
		\fntext[label2]{Department of Mechanical Engineering, Iowa State University, Ames, Iowa 50011, USA}
		\fntext[label3]{Ames National Laboratory, Division of Materials Science and Engineering, Ames, IA 50011, USA}
		\author{Raghunandan Pratoori\corref{cor1}\fnref{label1,label4}}
		\ead{rnp@iastate.edu}
		\author{Dmitry Popov\fnref{label4}}
		\author{Changyong Park\fnref{label4}}
		\author{Nenad Velisavljevic\fnref{label4}}
		\fntext[label4]{HPCAT, X-ray Science Division, Argonne National Laboratory, Lemont, IL, USA}
		\cortext[cor1]{Corresponding author}

	 	\begin{abstract}
		
 			Significant grain growth is observed during the high-pressure phase transformations (PTs) at room temperature within an hour for various materials~\citep{popov2019real,velisavljevic2011effects,cunningham2005crystal,gu1995crystal,mcmahon2000ba,li2018unusual,nelmes1993phase}. However, {no existing theory explains} this phenomenon since nanocrystals do not grow at room temperature even over a time span of  several years~\citep{edalati2022nanomaterials} because of slow diffusion. Here, we suggest a multistep mechanism for the grain growth during $\alpha\rightarrow\omega$ PT in Zr. Phase interfaces and grain boundaries (GBs) coincide and move together under the action of a combined thermodynamic driving forces.  Several intermediate steps for such motion are suggested and justified kinetically.  Nonhydrostatic stresses due to volume reduction in the growing $\omega$ grain promote continuous growth of the existing $\omega$ grain instead of a new nucleation at other GBs. In situ synchrotron Laue diffraction experiments confirm the main predictions of the theory. The suggested mechanism provides a new insight into synergistic interaction between PTs and microstructure evolution.
		\end{abstract}
	
		\begin{keyword}
				
				
				
		\end{keyword}
		
	\end{frontmatter}

 		Drastic grain growth  during the high-pressure PTs at room temperature at {an} hour time scale in zirconium~\citep{popov2019real,velisavljevic2011effects}, praseodymium~\citep{cunningham2005crystal}, cerium~\citep{gu1995crystal}, bismuth~\citep{mcmahon2000ba,li2018unusual},  InSb~\citep{nelmes1993phase}
 		{ and Ti (\Cref{S12})} challenges modern theory of grain growth involving thermally activated processes, like diffusion. Due to the slow diffusion at room temperature,  without PT,  nanograined $\omega$-Zr, for example (and many other metals with high melting temperature) after high-pressure torsion  does not show change in hardness and microstructure (which would be indicative of changing average grain size) for over ten years~\citep{edalati2022nanomaterials}. Increasing pressure does not change the hardness and grain size when materials are subjected to plastic straining ~\citep{Edalati-MT-10,Lin-Levitas-MRL-23}, and a hydrostatic loading reduces crystallite size for $\alpha$- and $\omega$-Zr rather than increasing them ~\citep{pandey2023effect}.

		Here, we suggest a mechanism and thermodynamic and kinetic theory for the grain growth during $\alpha\rightarrow\omega$ PT in Zr, consistent with existing and new experiments.	We will start theoretical interpreting the most recent experiment~\citep{popov2019real} on the nanograined $\alpha$-Zr when the nucleation and growth mechanism of $\omega$-Zr crystals has been clearly observed. Since the pressure in~\citep{popov2019real} (4.10 to 5.01 GPa)  is close to the phase equilibrium pressure for $\alpha$ and $\omega$ phase,
		$p_e=3.4$ GPa ~\cite{Pandey2020}, and below the most reported
		pressures for martensitic $\alpha\rightarrow\omega$ PT in Zr ~\cite{Vohra1978,olinger1973zirconium,velisavljevic2011effects,Pandey2020,pandey2023effect}, there is a small number of $\omega$ nuclei in separate grains located far away from each other, one nucleus per grain. We will focus on a single grain, in which an $\omega$ nucleus grows as a single crystal until it reaches the GBs and fills the entire grain. A hypothesis of this approach is that the phase and grain boundaries coincide (\Cref{Nucleus-1}) and move together under {the} action of combined thermodynamic driving force for PT and grain growth. This increases by several times the driving force for GB motion;  however, this still is not sufficient to provide any visible grain growth at room temperature (\Cref{sectionS8}). Thus, the new mechanism for grain growth is more related to the PT because PT occurs at a desirable rate at room temperature.

		Since high-angle GBs in nanograined materials are quite disordered, $\alpha\rightarrow\omega$ PT in Zr cannot be martensitic, but should occur through an intermediate disordered phase $\omega_d$. The {primary} condition is that the GB energy of the $\omega$ phase is smaller than that of the $\alpha$ phase, which promotes the nucleation of $\omega$-Zr in the $\alpha$ phase and is consistent with the absence of the reverse PT~\cite{Li2018a,Pandey2020,Lin-Levitas-MRL-23}. Incomplete $\omega$-Zr critical nucleus, $\omega_d$, heterogeneously nucleates in $\alpha$-Zr at the disordered high-angle GB, driven by {the} reduction in GB energy during the PT. Its structure is frozen along the path from $\alpha$ to $\omega$-Zr (\Cref{Nucleus-1}m), i.e., {it} contains some disordering, which relaxes to the disordered phase like in a GB. Thus $\omega_d\rightarrow$GB transformation occurs (\Cref{Nucleus-1}b-f) during lateral growth of this nucleus along the GB within $\alpha$-Zr, increasing the GB width. Nonhydrostatic stresses due to volume reduction in the growing $\omega$ grain increase the driving force and reduce the activation energy for such nucleation compared to other GBs. This {increase in driving force} explains why such nucleation does not routinely occur at other GBs, promoting the growth of the existing $\omega$ grain instead of a new nucleation.	Since deviation of the GB width from the thermodynamically equilibrium one is energetically penalized, the disordered material at the opposite side of the GB contacting $\omega$-Zr transforms to $\omega$-Zr, barrierlessly or via the critical nucleus (\Cref{Nucleus-1}g-j), restoring the GB width close to the thermodynamically equilibrium one. Thus, such multistage process leads to the shift of the combined $\alpha - \omega$ phase interface and GB. These two interfacial processes repeat themselves, producing continuous growth of $\omega$-Zr grain. The material exhibits nontrivial creativity in the synergistic interaction of two processes. This interaction changes the diffusional grain growth mechanism to the transformational one and the martensitic mechanism of PT to a reconstructive one via an intermediate disordered phase.
		Below, we will give thermodynamic and kinetic justification for each of the suggested steps.


		\begin{figure}[htp]
			\centering
			\resizebox{150mm}{!}{\includegraphics[trim={20mm 30mm 20mm 25mm},clip]{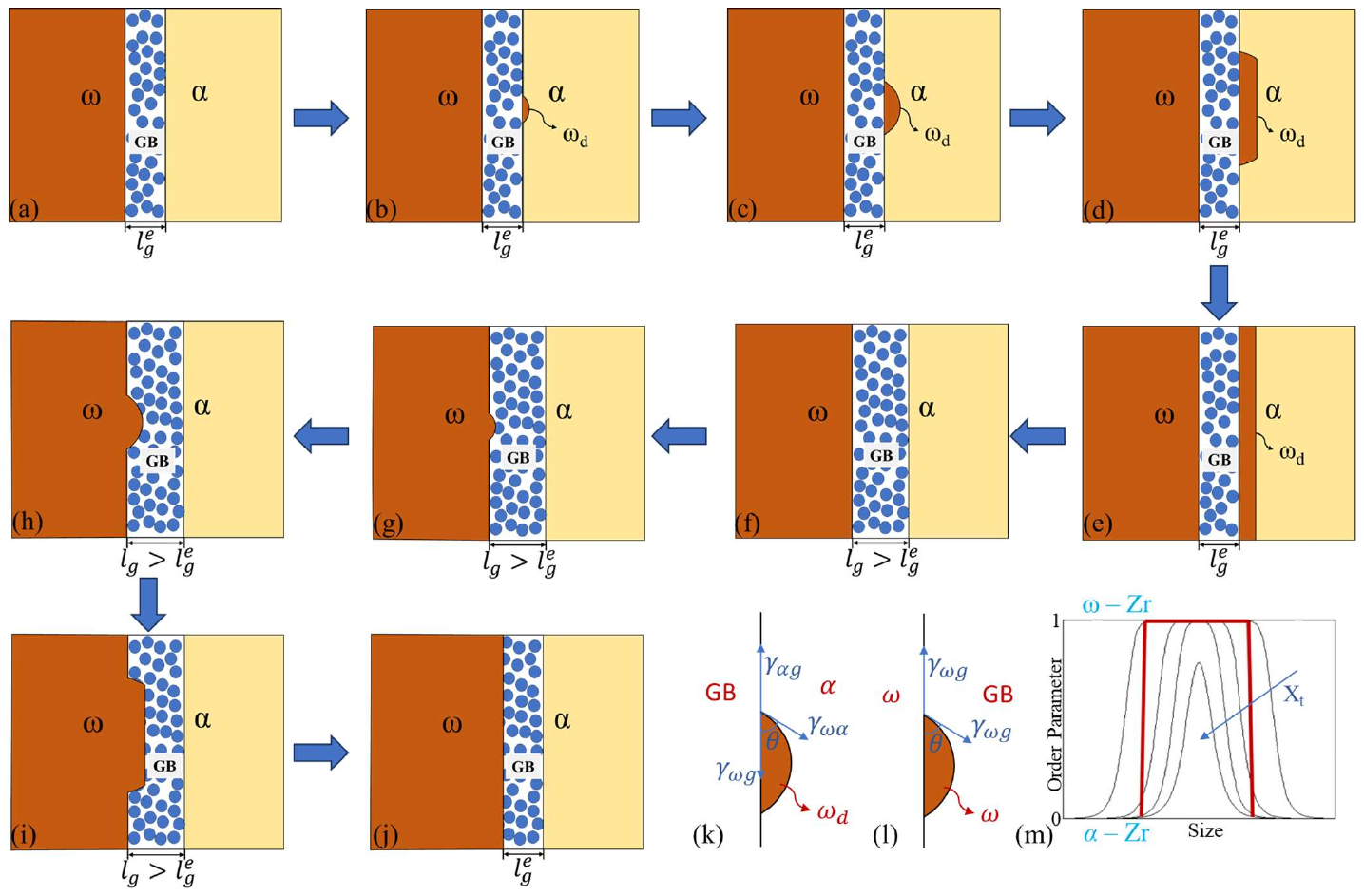}}
			\caption{\textbf{Schematic of the suggested sequence of the events for the suggested mechanism of the coupled grain boundary and phase interface propagation.} (a) Initial state of the $\alpha$ and $\omega$ phases divided by disordered GB with the thermodynamically equilibrium width $l_g^e$. (b)-(f)  Appearance of the critical nucleus  of the disordered $\omega_d$ phase at the combined grain and phase boundaries, its lateral  growth along the GB, and relaxation to the disordered GB with the width $l_g> l_g^e$. (g)-(j) Appearance of the critical nucleus  of the  $\omega$ phase within disordered GB and its lateral  growth along the GB leading to the equilibrium GB width. Steps (b)-(j)  results in the incremental shift of the combined phase and GB from $\omega$ to $\alpha$ phase. (k) Appearance of a critical nucleus of disordered $\omega$-Zr at the contact of the GB and $\alpha$-Zr. Vectors of the surface tension forces are shown.
			(l) Appearance of a critical nucleus of $\omega$-Zr within GB at the contact of the GB and $\omega$-Zr.
			(m)  Schematics of the order parameter profiles of the critical nucleus of $\omega$-Zr with increasing along the arrow $X_t$. Critical nuclei represents an intermediate (disordered) state between $\alpha$ and $\omega$ phases. Red box is the critical nucleus within sharp-interface approach.}
			\label{Nucleus-1}
		\end{figure}

		The thermodynamic driving force for a PT under pressure $p$ is
		\begin{equation}\label{XT1}
			X_t=-(p{-p_e)\varepsilon}_v,
		\end{equation}
		where  
		$\varepsilon_v=-0.0158$ is the volumetric transformation strain.  	PT in~\cite{popov2019real}  occurs in the pressure range from $4.10$ to $5.01$ GPa; we will use an average value $p=4.55$ GPa and evaluate
		\begin{equation}
			X_t = 18.17 MPa=18.17\times{10}^6\ J/m^3=\ 254.20 J/mol.
		\end{equation}
 		The energy of the spherical critical $\omega$-Zr nucleus within $\alpha$-Zr is~\cite{Porter1992}
		\begin{equation}\label{kinetic_g}
			G_{cr}=\frac{16 \pi \gamma_{\alpha\omega}^3}{3X_t^2} = 1.334\times{10}^{-16}J.
		\end{equation}
		The estimation of $\gamma_{\alpha\omega}=0.138\, J/m^2$ is given in \Cref{sectionS5}.
		Kinetic criterion for nucleation is~\cite{Porter1992}
		\begin{equation}\label{kinetic_nuc}
			G_{cr}\le(40-80)kT,
		\end{equation}
		where $k=1.381\times{10}^{-23}\ J/K$ is the Boltzmann constant, and $T$ is the absolute temperature. Taking $T=295 K$ and a factor of 40, we obtain $40kT=1.629\times{10}^{-19}\ J=0.0012G_{cr}$, i.e., homogeneous nucleation is impossible at room temperature.

		For heterogeneous $\omega$-Zr nucleation at the GB, the energy of the critical nucleus in the shape of a spherical cap (\Cref{Nucleus-1}k) is~\cite{Porter1992}
		\begin{align}\label{Het-nucl}
			G^{het}_{cr} = G_{cr}S(\theta);\quad
			S(\theta) = (2+\cos\theta)(1-\cos\theta)^2/4;\quad
			cos\theta = (\gamma_{\alpha g}-\gamma_{\omega g})/\gamma_{\alpha\omega},
		\end{align}
		where $\theta$ is the wetting angle. Determination of the shape factor $S\left(\theta\right)$ from the condition to make heterogeneous nucleation possible, i.e., $S\left(\theta\right)=0.0012$,  results in
		\begin{align}\label{theta_val}
			\cos{\theta}=0.959; \quad \theta={16.46}^\circ;\qquad
			\gamma_{\alpha g}-\gamma_{\omega g}=0.959\gamma_{\alpha\omega}=0.132\ J/m^2.
		\end{align}
		Obtained value of $\gamma_{\alpha g}-\gamma_{\omega g}$ is quite reasonable one (especially since our estimate for $\gamma_{\alpha\omega}$ in \Cref{sectionS5} is more than an order of magnitude larger than in  ~\cite{Yeddu2012}) and is expected in reality.
		A slightly supercritical nucleus will grow along the GB since it leads to a larger energy gain than the increase in its radius.  Note that for slightly larger $ \gamma_{\alpha g}-\gamma_{\omega g}= \gamma_{\alpha\omega} = 0.138\, J/m^2$, we have $\cos{\theta}=1$, $S\left(\theta\right)=0$, and barrierless GB-induced nucleation. The dependence of $cos\theta$ and $\theta$ versus pressure is shown in (\Cref{nhstresses}b). {Detailed discussion on the reduction in the GB energy during PT is given in \Cref{sectionSnew}.}
		
   		Note that more detailed treatments of nucleation at the surface or GB with the phase-field approach~\cite{Levitas2011,Levitas2014,Basak2018,Basak2021,Levitas2018,Momeni2015}, especially related to the formation of the critical nucleus~\cite{Levitas2014,Momeni2015}, provide much more detail. Surface-induced critical nucleus represents an intermediate variable structure $\omega_d$ between the parent and product phases, significantly reducing its energy.  The complex atomic pathway between $\alpha$ and $\omega$ phases of Zr implies that the intermediate structure may be similar to the disordered structure of the GB and will be transformed to it during nucleation and lateral growth along the GB (\Cref{Nucleus-1}(m)). In fact, one of the main mechanisms of amorphization under pressure is related to the arrest of intermediate phases between two crystal phases due to slow kinetics. Thus, we arrived at the following multistage mechanism of the elemental GB advance (\Cref{Nucleus-1} (a)-(f)):
	   	\begin{center}
	   	    $\alpha\ \rightarrow$ Critical nucleus of $\omega_d\rightarrow$ Lateral growth along the GB and transformation to GB.
	   	\end{center}

  	The question arises: why a typical heterogeneous nucleation of $\omega$ phase without transformation to a disordered GB does not routinely occur at any GB, leading to multiple nuclei and grains of $\omega$-Zr? We will show that the effect of internal stresses due to volume reduction during $\alpha-\omega$ PT, especially nonhydrostatic stresses, explains the reasons for the growth of a small number of $\omega$ grains to large sizes.

  	\begin{figure}[htp]
			\centering
        \resizebox{150mm}{!}{\includegraphics[trim={20mm 115mm 70mm 50mm},clip]{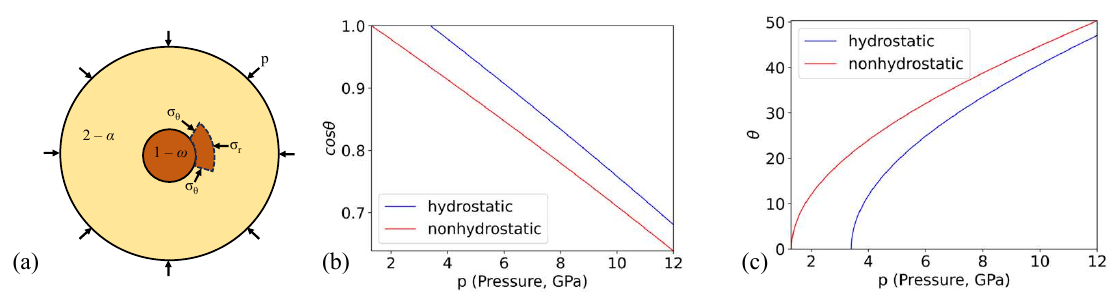}}
			\caption{\textbf{Effect of  nonhydrostatic stresses caused by a spherical $\omega$ grain on the heterogeneous nucleation of  $\omega$ phase.} (a) Schematic for determination of stresses in the growing spherical $\omega$ grain (phase 1) and surrounding $\alpha$ phase (phase 2). Nonhydrostatic stresses are shown at the element in the shape of the part of the spherical ring at the combined phase interface and grain boundary, which is also considered as the critical nucleus of the $\omega$ phase. Effect of pressure on (b) $\cos \theta$  and (c) wetting angle $\theta$ for hydrostatic and nonhydrostatic loading.}
			\label{nhstresses}
		\end{figure}

		Reduction in volume in the growing $\omega$ grain during $\alpha-\omega$ PT leads to nonhydrostatic internal stresses in the surrounding $\alpha$ phase, which significantly promotes PT. For the spherical $\omega$ grain with the radius $R$ within infinite $\alpha$ space under the action of the external pressure $p$ (\Cref{nhstresses}a),  we obtain  for the radial and azimuthal stresses in the $\alpha$ phase at the boundary with the growing $\omega$ phase grains $r=R$ and for the pressure in the $\omega$ phase, $p_1$  (\Cref{sgr20})
		\begin{align}\label{sgr20}
			&\sigma_{r2}(R) = \sigma_{r1} = -p_1 = -\frac{K_1(p(3K_2 + 4\mu_2) + 4\mu_2K_2\varepsilon_v)}{K_2(4\mu_2 + 3K_1)}; \\
			&\sigma_{\theta 2}(R) = \frac{p(2K_1\mu_2-3K_1K_2-6K_2\mu_2)+2\mu_2K_1K_2\varepsilon_v}{K_2(4\mu_2+3K_1)}. \nonumber
		\end{align}
		Using known pressure dependence of the bulk and shear moduli for Zr phases~\cite{pandey2023effect,Fisher1970,Liu2007}   (\Cref{modulip}) and applying $p=4.55$ GPa, we obtain \Cref{moduli-4}
		and
		\begin{align}\label{calc_stress}
			\sigma_{r2}\left(R\right)=\sigma_{r1}=-p_1=-4.118\ GPa;\qquad
			\sigma_{\theta2}\left(R\right)=-4.766\ GPa;\qquad
			p_2=4.55\ GPa.
		\end{align}
		Since $\sigma_{r1}-\sigma_{\theta2}=0.648$ GPa is smaller than the yield strength of $\alpha$ and $\omega$ phases even at zero pressure~\cite{Pandey2020}, neglecting plastic strains is justified.
		The deviatoric stresses at the GB promote heterogeneous nucleation of the critical $\omega$-Zr nucleus at the GB, which is shown as part of the spherical ring in \Cref{nhstresses}a.  Since the nucleus is of a few nm size, the curvature of the GB is not important, and we return to the schematic in \Cref{Nucleus-1}. We will neglect the change in stresses in \Cref{calc_stress} during the appearance of a small critical nucleus. The normal components of the  transformation strain tensor for $\alpha- \omega$ PT in Zr are~\cite{Yeddu2012}
		$\tilde{\fvep}_{t}=\{	-0.0318, 0.038,  -0.022 \}$.
		The maximum driving force for PT will be when two compressive components of $\tilde{\fvep}_{t}$ are oriented along the azimuthal stress, and the tensile component is along the radial direction. Then
		\begin{align}\label{Xt-nonhyd}
			X_t=\sigma_{\theta2}\left(\varepsilon_{t1}+\varepsilon_{t3}\right)+\sigma_{r1}0\varepsilon_{t2}+ p_e\varepsilon_v=\sigma_{\theta2}\left(-0.0318-0.022\right)+\sigma_{r1}0.038-0.0537=46.21\times{10}^6\ J/m^3.
		\end{align}
		This driving force is 2.541 times larger than that under the same hydrostatic pressure of 4.55 GPa, which according to \Cref{kinetic_g} leads to $(2.541)^2=6.454$ time reduction in the activation energy, down to $G_{cr}= 2.062\times{10}^{-17}J$.    This increases the required $S\left(\theta\right)$ by a factor of 6.454 to $S\left(\theta\right)=0.0080$ leading to
		\begin{align}\label{new_theta}
			\cos{\theta}=0.895; \qquad \theta={26.54}^\circ; \qquad
			\gamma_{\alpha g}-\gamma_{\omega g}=0.895\gamma_{\alpha\omega}=0.123\ J/m^2.
		\end{align}
		Note that while nucleation occurs at a place with optimally oriented  $\alpha$ crystal, propagation along a curved interface
		does not require this condition, which is consistent with isotropic grain growth in experiments~\citep{popov2019real}.
		Thus, including the effect of nonhydrostatic stresses weakens requirements for reducing GB energy during $\alpha-\omega$ PT and makes our scenario more plausible. More importantly, this explains why such a nucleation does not routinely occur at other GBs, promoting the growth of the $\omega$ grain instead of multiple nucleations. The dependence of $\cos{\theta}$ and $\theta$ on pressure is shown in (\Cref{nhstresses}c).

		To complete the elemental event of the GB displacement, the GB width should be reduced to the equilibrium value \Cref{leq} by PT of the disordered Zr within the GB side, which is in contact with $\omega$-Zr, to $\omega$-Zr   (\Cref{Nucleus-1} (g)-(j)). Let us consider the appearance of a critical $\omega$-Zr nucleus in the shape of a spherical cap (\Cref{Nucleus-1}l). Since after its appearance, its flat part and corresponding surface tension $\gamma_{\omega g}$ disappear, the only way to satisfy the mechanical equilibrium due to surface tensions is to set the wetting angle $\theta=0$. Then $S\left(\theta\right)=0$, and the energy of the critical nucleus is zero as well, i.e., barrierless nucleation occurs. This means that in the continuum treatment plane, GB-$\omega$ phase interface can move barrierlessly under any positive thermodynamic driving force. However, this is unlikely  because the discrete atomic structure would cause a barrier to the motion of the plane GB-$\omega$ phase interface. That is why we consider the appearance of a finite $\omega$-Zr nucleus within disordered Zr in the shape of a spherical cap (\Cref{Nucleus-1}l) with some finite wetting angle $\theta$ (to be found) and size exceeding a few crystal cells and check the satisfaction of kinetic criterion \Cref{kinetic_nuc}. To be safe, we neglect the increase in the driving force for nucleation due to nonhydrostatic stresses (they may relax in a disordered GB) and deviation of the GB width from the equilibrium value.

		Change in the Gibbs energy due to the appearance of such a nucleus is~\cite{Porter1992}
		\begin{align}\label{dG}\c{}
			\Delta G=&-X_t\pi\ R^3\left(2+\cos{\theta}\right)\left(1-\cos{\theta}\right)^2/3+\gamma_{\omega g}\pi\ R^2[2\left(1-\cos{\theta}\right)-\sin^2\theta]=[-X_t\pi R^3+\gamma_{\omega g}\pi R^2]\theta^4/4,
		\end{align}
		where  the Taylor series in $\theta$ up to the fifth degree was used.
		Then the critical radius and activation energy are determined by maximizing $\Delta G$ with respect to $R$:
		\begin{align}
			R_{cr}=\frac{{2\gamma}_{\omega g}}{3X_t}=5.36\ nm;\qquad
			G_{cr}= \frac{\pi\gamma_{\omega g}^3}{27X_t^2}\theta^4=2.924\times10^{-18}\theta^4 J,
		\end{align}
where $\gamma_{\omega g}=0.389\ J/m^2$ from \Cref{sectionS6} was used.
		From the condition $G_{cr}=40kT$,  we obtain $\theta=0.486={27.84}^\circ$. Note that the difference between the exact functions of $\theta$ in \Cref{dG} and the Taylor series for this $\theta$ does not exceed 8\%. The height $h_c$ of the spherical cap with respect to the initial plane interface and its radius $R_c$ within an interface (\Cref{Nucleus-1}l) are
		$	h_c=R_{cr}\left(1-\cos{\theta}\right)=0.62\ nm$;
		$			R_c=R_{cr}\sin{\theta}=2.50\ nm.$
		Since 	these sizes exceed lattice parameters $a=0.5039$ nm and $c=0.3136$ nm of  $\omega$-Zr~\cite{Vohra1978,olinger1973zirconium}, the nucleus contains several lattice cells. After the fluctuational appearance of such a mechanically and thermodynamically nonequilibrium $\omega$-Zr nucleus, it will spread laterally along the GB-$\omega$ interface, completing the elemental step of the GB motion in \Cref{Nucleus-1} (g)-(j). This concludes the thermodynamic and kinetic justification of the suggested mechanism.

		The next question is: while at the initial stage in~\citep{popov2019real}, grains are not in contact, what happens at a later stage of PT when grains come into contact? Will  $\omega$-Zr grain grow at the expense of other $\omega$-Zr grains?
		While results in~\citep{popov2019real} were reported for the pressure increased from 4.1 to 5.0 GPa increased during 66 minutes, here we post-processed the data during further pressure increase to 8.9 GPa between 66 and  311 minutes from the onset of grain enlargement process (\Cref{9.1}). It was found  that nucleation and enlargement of $\omega$-Zr crystals at 66 minutes after the onset was much slower compared to the period of time right after the onset, which is consistent with traditional PT kinetics. Increase of the overall diffraction intensities in this period of time was much smaller than previously (\Cref{Fig3}a,b, Movie S1). Growth of selected single-crystals of $\omega$-Zr has been observed directly by mapping of their reflections. For this purpose, in total 10 $\omega$-Zr
 crystals, distributed within the entire sample, have been identified first by indexing of their Laue patterns (\Cref{Fig3}c, Movies S2 and S3).
		Out of the selected 10 crystals, 7 did not exhibit notable enlargement starting from 66 minutes after the onset (\Cref{Fig4}a,b, movies S2 and S3).
		Other 3 crystals after 66 minutes exhibited essential growth in only one direction, while before 66 minutes all 10 crystals grew isotropically (\Cref{Fig4}c,d).
		This evolution can be explained if  crystals are in the contact and 3 of them had sufficiently high driving force to grow in the direction where there was a place to grow, but the rest could not grow because of the small space around and the small driving force due to developed stresses. It is also proper to recall that the $\omega$-Zr crystals may have very irregular shape as it was identified by their high resolution mapping (\Cref{9.2}, \Cref{S4}). This observation also may be attributed to limited space to grow at the final stage of this PT. Stresses cannot be evaluated analytically because
		of a complex interaction of stress fields of different grains.
		Also, no notable shrinking of 10 selected $\omega$-Zr crystals was observed  within the entire studied pressure range, which excludes $\omega$-Zr grain growth at the expense of other $\omega$-Zr grains. Likewise, in case of growth of $\omega$-Zr crystals in expense of other $\omega$-grains the overall number of Laue spots would exhibit notable drop while their intensities would notably increase during this process. This was not the case even when the grains came into contact (\Cref{Fig3}a,b). Thus, $\omega$-grain growth occurs during PT only.

		In another experiment (\Cref{9.3}), the grain enlargement process across the $\alpha - \omega$ transition in Zr was observed with much better spatial resolution than in the experiment above and in~\citep{popov2019real} but focusing on one $\omega$-Zr crystal {(Movie S4)}. Again, no shrinking of the studied crystal was observed but only enlargement and stabilization. The studied crystal gradually enlarged within 2 hours and 32 minutes and remained stable during further compression.

		The $\omega$-grains exhibit high density of crystal lattice defects indicated by "streaky", irregular shape of their Laue spots {(\Cref{Fig3}a,b, \Cref{Fig4})}. Sub-grain boundaries in an $\omega$-grain have been also observed using X-ray diffraction with sub-micron beam~\citep{popov2019real}. During this experiment, due to much better spatial resolution, some release of the $\omega$ -Zr crystal from defects was observed after the crystal size stabilized. Namely, some Laue reflections became somewhat sharper, indicating that the dislocation density and lattice microstrain became smaller {(\Cref{S6})}. This agrees with our mechanism because the crystallization of the $\omega$ -Zr from the disordered grain boundary is a highly nonequilibrium process, leading to numerous defects that relax/reorganized in time. Despite current and reported in~\citet{popov2019real} results have been obtained with a similar average pressure rate of about 0.8 GPa/hour, enlargement of $\omega$ -Zr crystals in~\citep{popov2019real} mainly was completed in a much shorter time of 1 hour and 6 minutes. This difference is probably caused by different distributions of residual strain, dislocations, grain sizes, and intergranular structure in the original samples of $\alpha$ -Zr because scraping the samples from the bulk sample surface was done manually without taking care of the reproducibility of their microstructure.

		To verify the role of nonhydrostatic stresses induced by growing $\omega$-Zr grain, another experiment was performed for $p\leq p_e=3.4\, GPa$ (\Cref{9.4}). Sample of $\alpha$-Zr was compressed up to 3.24 GPa. No Laue spots, indicative of $\omega$-Zr crystals, were observed during and right after compression. The sample was kept without changes after compression and after one year and ten months diffraction spots from $\omega$-Zr grains clearly dominated over the rings from $\alpha$-Zr observed with monochromatic beam. { In the independent experiment, $\alpha$-Zr was kept in DAC at pressure $\sim 3\, GPa$, and the following measurements were performed in 3 months.
Change in pressure was within an experimental error, and the entire sample was transformed to large-grain $\omega$ phase.}
		As it follows from \Cref{nhstresses}b,c, under hydrostatic loading, PT is impossible even for $\cos{\theta}=1$ since $X_t\leq 0$ in \Cref{XT1}.
		When the effect of nonhydrostatic stresses due to growing $\omega$-Zr grain is included, $X_t\geq 0$ in \Cref{Xt-nonhyd} for  $1.31\, GPa \leq p\leq p_e $ and  $\omega$-Zr grain growth is thermodynamically possible. The only remaining question on the appearance of the initial embryos  of each $\omega$-Zr grain under initial hydrostatic pressure can be resolved  as follows. If in some places of the initial
		GB in $\alpha$-Zr
		$\Delta\gamma_{\omega} = \gamma_{\alpha g}-\gamma_{\omega g}- \gamma_{\alpha\omega} > 0$ (i.e., nucleation of $\omega$-Zr leads to the reduction in the total surface energy, the critical nucleus is not necessary, and  equilibrium $\theta$ in \Cref{Het-nucl} does not exist),
		barrierless nucleation of the equilibrium nm-thin layer of $\omega$-Zr of the width
		\begin{equation}\label{leqomeg}
			l_\omega^e = \delta_\omega \ln\left(-\frac{\Delta\gamma_\omega}{X_t \delta_\omega}\right); \quad -\frac{\Delta\gamma_\omega}{X_t\delta_\omega}\geq 1,
		\end{equation}
		may occur under hydrostatic loading even at  $X_t\leq 0$ (see \Cref{leqom});  here,   $\delta_\omega$ is a characteristic interaction length between $\alpha$-GB and $\omega$-GB interfaces.
		 By approaching thermodynamic equilibrium $X_t\rightarrow 0$, width
		$l_\omega^e \rightarrow \infty$, i.e., $\omega$ phase theoretically spreads over the entire grain if the GB length at which $\Delta\gamma_{\omega}>0$ and entire grain are infinite.
		For nanometer grain size, the same process is possible  at much smaller $X_t\leq 0$ ~\cite{Couchman-77,Levitas2014,levitas-samani-11} (because small grains further promote GB-induced PT), but small
		GB area at which $\Delta\gamma_{\omega}>0$ suppresses PT. If such a combination produces an initial $\omega$-Zr nucleus close to the spherical shape, it will produce nonhydrostatic stresses  in \Cref{sgr20} and corresponding driving force $X_t$ in  \Cref{Xt-nonhyd}, which will
		promote further growth.

		\begin{figure}[htp]
			\centering
        \resizebox{150mm}{!}{\includegraphics{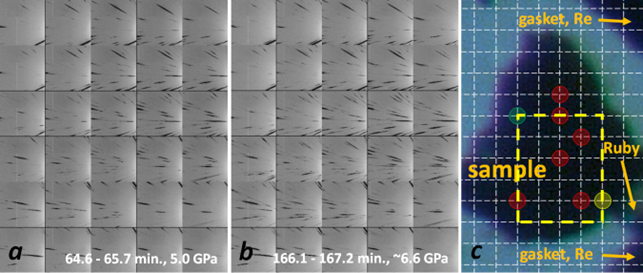}}
			\caption{\textbf{Maps of Laue patterns and optical image of the sample.} (a) and (b) Composite frames of Laue diffraction patterns and time intervals when the data were collected, starting from the onset of the $\omega$-Zr grain enlargement process. Pressure in (a) was measured using Ruby system~\citep{mao1986calibration} soon after the data were collected while pressure in (b) was estimated based on an approximation (Supplementary Material). Intensity scaling in (a) and (b) is the same in Fit2d notation~\citep{hammersley1996two}. All obtained frames are compiled as a movie (Movie S1). (c) Optical image of the sample in a diamond anvil cell. White dotted grid denotes projection of the 2D scans. Step size was 5 $\mu$m. Yellow and green circles denote scan positions closest to centroids of crystals presented in Movies S2 and S3 respectively, while red circles denote scan positions closest to centroids of other identified $\omega$-Zr crystals. Yellow dotted rectangle outlines area presented in (a) and (b). }
			\label{Fig3}
		\end{figure}

		\begin{figure}[htp]
			\centering
        \resizebox{100mm}{!}{\includegraphics{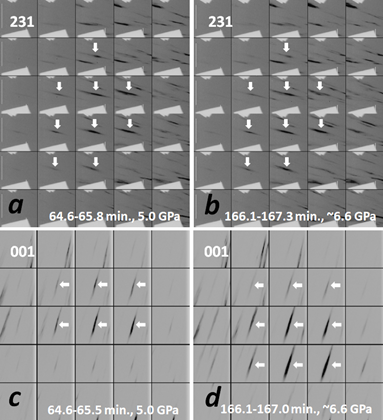}}
			\caption{\textbf{Maps of Laue reflections from $\omega$-Zr crystals.} (a) and (b) Maps of Laue reflections from $\omega$-Zr crystal presented in Movie S3  and (c) and (d) from another $\omega$-Zr crystal. White arrows help to distinguish these reflections from other Laue spots. Time intervals when the data were collected are specified starting from the onset of the $\omega$-Zr grain enlargement process. Intensity scaling for each crystal is the same in Fit2d notation~\citep{hammersley1996two}.}
			\label{Fig4}
		\end{figure}

		In ~\citep{velisavljevic2011effects}, for initially nanograined Zr with high impurities content, $\alpha\rightarrow\omega$ PT starts at 11.6 GPa, and then the temperature was increased up to 1279 K.
		However, grain growth was not observed. Based on our mechanism, increased PT pressure can be caused by the
		increased GB energy of $\omega$-Zr and $\cos{\theta}$ (\Cref{nhstresses}b,c), reducing the GB's role as a nucleating defect. Then nucleation
		may occur at dislocations, twins, or other defects, and our mechanism does not work. Also, due to large $X_t$, nucleation may start initially at GB at many places of  the same $\alpha$ grain and in many $\alpha$ grains, leading to a large number of  $\omega$-grains and continuous diffraction rings.
		The lack of significant grain growth at 1279 K confirms that traditional mechanisms do not work on the experimental time scale of even up to hours.
		These high-pressure results contrast our experiments above, in which large grains were observed at a pressure of  8.9 GPa.
		However, here, the major part of the  PT and grain growth occurred below 5 GPa and only partially continued till  8.9 GPa, leaving little space for
		multiple new grain nucleation.  In ~\citep{velisavljevic2011effects}, PT started at 11.6 GPa and with increasing temperature, leading to
		numerous nucleation events from the beginning.

		For plastic strain-induced $\alpha\rightarrow\omega$ PT, grain growth is not observed ~\citep{Pandey2020,Lin-Levitas-MRL-23}. This is consistent with the suggested in ~\citep{levitas-prb-04,Levitas-MT-19}
		mechanism of strain-induced PTs based on nucleation at the tip of dislocation pileup instead of GB promoted nucleation discussed here.

		\textbf{Method:}
		The grain enlargement process of $\omega$-Zr has been observed in-situ and en-operando, using the Laue diffraction technique, with experimental setups which have been available at 16BMB beamline of the Advanced Photon Source (APS) ~\citep{popov2019mechanisms,popov2015high}. These setups have been optimized explicitly for Laue measurements in diamond anvil cells (DAC). Details of the previous experiment are reported in~\citet{popov2019real}. The experiment reported here has been conducted with the setup provided much more stable X-ray beam profiles and sample positioning~\citep{popov2019mechanisms} (\Cref{9.3}). Sample of $\alpha$-Zr for compression was prepared the same way as for the previous experiment~\citep{popov2019real}. Ne was used as a pressure-transmitting medium providing hydrostatic compression. Laue diffraction measurements have been conducted using the transmission-mode geometry with the area detector tilted vertically by 30$^\circ$. X-ray beam having an energy range of 10-70 keV was focused with KB-mirrors down 2.4 $\mu$m horizontally and 2.6 $\mu$m vertically at half width maximum of the beam profiles presented in Supplementary Materials. A series of 2D scans have been collected on the sample across compression from 0.9 to 11.51 GPa with a step of 0.5 $\mu$m.
		Data analysis included indexing and mapping Laue reflections~\citep{popov2019real,popov2019mechanisms}. The indexation routine was ad-hoc modified to account for sample rotation caused by the shrink of the gasket (\Cref{9.1.2}). Another sample of $\alpha$-Zr was prepared the same way, compressed in a DAC up to 3.24 GPa using Ne as pressure transmitting medium, and kept without changes. Laue data was collected on the sample during compression using an experimental setup available at the microdiffraction beamline 12.3.2 of Advanced Light Source~\citep{kunz2009dedicated}. After one year and ten months, this sample was studied with monochromatic beam diffraction technique at 16BMD beamline at the APS~\citep{park2015new}. All the experiment details are given in \Cref{9.4}.

\noindent {\bf Acknowledgement}
\par VIL and RP acknowledge the support of NSF  (DMR-2246991 and CMMI-1943710) and Iowa State University (Vance Coffman Faculty Chair  Professorship and Murray Harpole Chair in Engineering). The simulations were performed at  Extreme Science and Engineering Discovery Environment (XSEDE), allocation TG-MSS170015. Portions of this work were performed at HPCAT (Sector 16), Advanced Photon Source (APS), Argonne National Laboratory. HPCAT operations are supported by DOE-NNSA’s Office of Experimental Sciences.  The Advanced Photon Source is a U.S. Department of Energy (DOE) Office of Science User Facility operated for the DOE Office of Science by Argonne National Laboratory under Contract No. DE-AC02-06CH11357. This research also used BL12.3.2 which is a resource of the Advanced Light Source, which is a DOE Office and Science User Facility under contract no. DE-AC02-05CH11231.

\newpage
\bibliography{ref_updated}

\pagebreak
\begin{frontmatter}
		
		
		\setcounter{author}{0}
		\title{ -- Supplementary material}
		

\vspace{-25mm}

	\end{frontmatter}
\setcounter{equation}{0}
\setcounter{figure}{0}
\setcounter{table}{0}
\setcounter{page}{1}
\makeatletter
\renewcommand{\theequation}{S\arabic{equation}}
\renewcommand{\thefigure}{S\arabic{figure}}
\renewcommand{\thesection}{S\arabic{section}}

\section{Driving force for \texorpdfstring{${\alpha}\rightarrow{\omega}$}{a-w} PT in Zr}

		In the experiment~\cite{popov2019real}, PT occurs in the pressure range from $4.10$ to $5.01$ GPa; we will use an average value for pressure $p=4.55$ GPa for numerical estimates; the phase equilibrium pressure is $p_e=3.4$ GPa~\cite{Pandey2020}. The thermodynamic driving force for a PT is
		\begin{equation}\label{drivingF}
			X_t = -p\varepsilon_v - \Delta\psi,
		\end{equation}
		where $\varepsilon_v$ is the volumetric transformation strain, and ${\Delta}\psi$ is the jump in thermal energy; the contribution due to the jump in elastic moduli is neglected. Phase equilibrium pressure is determined from the condition
		\begin{equation}
			X_t = -p_e\varepsilon_v-{\Delta}\psi=0.
		\end{equation}
		Then
		\begin{equation}\label{XT}
			X_t=-(p{-p_e)\varepsilon}_v.
		\end{equation}
		Since for $\alpha\rightarrow\omega$ PT in Zr $\varepsilon_v=-0.0158$, then at $p=4.55$ GPa
		\begin{equation}
			X_t = 18.17 MPa=18.17\times{10}^6\ J/m^3=\ 254.20 J/mol.
		\end{equation}
We used the atomic volume of Zr of  $13.99\times10^{-6} \,  m^3/mol$.

	    \section{Effect of nonhydrostatic internal stresses in the growing \texorpdfstring{${\omega}$}{w} grain on thermodynamics and kinetics of \texorpdfstring{${\alpha}-{\omega}$}{a-w} PT at the GB}

    	The question arises: why such a heterogeneous nucleation of $\omega$ phase but without transformation to a disordered GB does not routinely occur at any GB, leading to multiple nuclei and grains in $\omega$-Zr? We will show that the effect of internal stresses due to volume reduction during $\alpha-\omega$ PT, especially nonhydrostatic stresses, explains the reasons for the growth of a small number of $\omega$ grains to large sizes.


    	Reduction in volume in the growing $\omega$ grain during $\alpha-\omega$ PT leads to nonhydrostatic internal stresses in the surrounding $\alpha$ phase, which significantly promotes PT. Let us consider spherical $\omega$ grain with the radius $R$ within infinite $\alpha$ space under the action of the external pressure $p$ (\Cref{nhstresses}a). Radial distribution of the radial $\sigma_{ri}$ and azimuthal $\sigma_{\theta i}$ stresses and pressure $p_i$ in phases $i=1$ ($\omega$ phase) and $i=2$ ($\alpha$ phase) are defined as~\cite{hill1950mathematical}
    	\begin{align}
    		&\sigma_{ri} = -\frac{4\mu_iB_i}{r^3} + 3K_i\left(A_i - \varepsilon_v/3\right); \label{stresses1}\\
    		&\sigma_{\theta i} = \frac{2\mu_iB_i}{r^3} + 3K_i\left(A_i - \varepsilon_v/3\right); \label{stresses2}\\
    		&p_i = -\left(\sigma_{ri} + 2\sigma_{\theta i}\right)/3 = 3K_i(A_i - \varepsilon_v/3)\label{stresses3}.
    	\end{align}
    	Here, $r$ is the position vector, $\mu_i$ and $K_i$ are the shear and bulk moduli of phase $i$, and four constants $A_i$ and $B_i$ are determined from the boundary conditions. Preliminary analyses of \Cref{stresses1,stresses2,stresses3} shows the following. Since at $r=0$ stresses should be finite, then $B_1=0$, stresses in $\omega$ phase are homogeneous, hydrostatic and equal to the negative pressure $-p_1$. While stresses in the $\alpha$ phase depend on $r$, pressure does not. For $r\rightarrow\infty$ we obtain
    	\begin{equation}
    		\sigma_{r2} = \sigma_{\theta 2} = -p_2 = 3K_2\left(A_2 - \frac{\varepsilon_v}{3}\right) = -p,
    	\end{equation}
		where the last equality follows from the boundary condition and determines $A_2$. Thus, pressure in the $\alpha$ phase is constant and equal to the applied pressure. Constant pressure in the $\omega$ phase for compressive (negative) $\varepsilon_v$ is supposed to be smaller than $p$, which will be confirmed below. Deviatoric stresses in the $\alpha$ phase (the first terms in \Cref{stresses1,stresses2}) are maximum at the interface ($r=R$) and decay with increasing $r$ down to zero at infinity.

		To determine $A_1$ and $B_2$, we apply at the boundary $r=R$ the continuity of the radial displacements and radial stresses~\cite{hill1950mathematical}. Then we obtain for the pressure in the $\omega$ phase
		\begin{equation}\label{sgr1}
			\sigma_{r1} = -p_1 = -\frac{K_1(p(3K_2 + 4\mu_2) + 4\mu_2K_2\varepsilon_v)}{K_2(4\mu_2 + 3K_1)},
		\end{equation}
		and for stresses in the $\alpha$ phase at the boundary with the growing $\omega$ phase grains $r=R$
		\begin{align}\label{sgr2}
			&\sigma_{r2}(R) = \sigma_{r1} = -p_1 = -\frac{K_1(p(3K_2 + 4\mu_2) + 4\mu_2K_2\varepsilon_v)}{K_2(4\mu_2 + 3K_1)}; \\
			&\sigma_{\theta 2}(R) = \frac{p(2K_1\mu_2-3K_1K_2-6K_2\mu_2)+2\mu_2K_1K_2\varepsilon_v}{K_2(4\mu_2+3K_1)}.
		\end{align}

		Using known pressure dependence of the bulk and shear moduli for Zr phases~\cite{pandey2023effect,Fisher1970,Liu2007}
		\begin{align}\label{modulip}
			&K_1 = 102.4+2.93p\ GPa;\qquad
			K_2=93.55+3p=107.20\ GPa;\nonumber \\
			&\mu_1=45.1+0.6p=47.83\ GPa; \qquad
			\mu_2=36.13+0.02p=36.22\ GPa,
\end{align}
and applying $p=4.55$ GPa, we obtain
		\begin{align}\label{moduli-4}
			&K_1 = 102.4+2.93\times4.55=115.73\ GPa;\qquad
			K_2=93.55+3\times4.55=107.20\ GPa;\nonumber \\
			&\mu_1=45.1+0.6\times4.55=47.83\ GPa; \qquad
			\mu_2=36.13+0.02\times4.55=36.22\ GPa,
\end{align}
and
\begin{align}\label{calc_stress-0}
			\sigma_{r2}\left(R\right)=\sigma_{r1}=-p_1=-4.118\ GPa;\qquad
			\sigma_{\theta2}\left(R\right)=-4.766\ GPa;\qquad
			p_2=4.55\ GPa.
		\end{align}
		Since $\sigma_{r1}-\sigma_{\theta2}=0.648$ GPa is smaller than the yield strength of $\alpha$ and $\omega$ phases even at zero pressure~\cite{Pandey2020}, neglecting plastic strains is justified. Thus, pressure in the growing $\omega$ grain is lower than the applied pressure, and pressure at the GB in the $\alpha$ phase is equal to the applied pressure, but there are also deviatoric stresses, which increase the driving force for PT. Consequently, the thermodynamic driving force for the appearance of an isolated spherical nucleus of $\omega$ phase should be calculated based on the averaged pressure at the beginning (4.55 GPa) and completing (4.118 GPa)  the PT ~\cite{Levitas1997,Levitas2021}, i.e., $p_{av}=4.334$ GPa
		\begin{equation}
			X_t = -(p_{av}-p_e)\varepsilon_v = 14.75\ MPa=14.75\times{10}^6\ J/m^3.
		\end{equation}
		This reduction in driving force increases the activation energy to $2.023959\times{10}^{-16}\, J$, which, however, does not change the main conclusion that homogeneous nucleation of $\omega$-Zr is impossible.

	\section{Thermodynamically equilibrium width of grain and phase boundaries}

		We will consider a GB coinciding with the $\alpha - \omega$ phase interface and having the same width. Let us start with the $\alpha - \omega$ phase interface in the single crystal. Then, in a thought experiment, let us rotate the crystal lattice of $\alpha$ phase, producing a GB coinciding with the $\alpha - \omega$ phase interface. For a small misorientation angle, rotation can be reproduced by an array of parallel equally-spaced dislocations~\cite{Porter1992}, which determine and excess energy of the GB. With increasing misorientation, spacing between dislocations reduces, and their core overlap, and for misorientation greater than $10^\circ-15^\circ$, it is problematic to distinguish separate dislocations. Thus, the GB represents a disordered material. Since nanograined material possesses mostly high-angle GBs, we will consider them a disordered phase with the finite width $l_g$, and determine a thermodynamically equilibrium width $l_g^e$. The energy of the finite-width GB per unit area, $\Gamma_{gb}$, versus the GB width $l_g$  is~\cite{Luo2008,Levitas2012,Momeni2014}
		\begin{align}\label{gamma_gb}
			&\Gamma_{gb} = \gamma_{\alpha g} + \gamma_{\omega g} + (G_d - G_a)l_g + \Delta\gamma_g exp\left(-\frac{l_g}{\delta_g}\right); \nonumber \\
			&G_d > G_\alpha; \quad \Delta\gamma_g = \gamma_0 - \gamma_{\alpha g} - \gamma_{\omega g} > 0.
		\end{align}

		Here, $G_d$ and $G_\alpha$ are the Gibbs potentials per unit volume of the disordered phase and $\alpha$-Zr, $\gamma_0$, $\gamma_{\alpha g}$, and $\gamma_{\omega g}$ are the energy per unit area of the zero-width GB (which represents sharp $\alpha -\omega$  phase interface), $\alpha$-GB, and $\omega$-GB interfaces, respectively; $\Delta\gamma_g$
is the difference  in the energy between initial zero-width GB and final  $\alpha$-GB and $\omega$-GB interfaces.
 The exponential term describes the short-range repulsion between $\alpha$-GB and $\omega$-GB interfaces, which produces the finite width of the disordered phase and GB; $\delta_g$ is a characteristic interaction length. While various other types of repulsions between interfaces (e.g., the long-range dispersion interaction) are possible~\cite{Luo2008,Levitas2012,Momeni2014}, they lead to the same qualitative results. It is natural that for the disordered phase $G_d > G_\alpha$  and $G_d > G_\omega$, otherwise, the entire volume will tend to transform into a disordered phase. Since $G_\alpha > G_\omega$  and $\alpha$-Zr transforms to $\omega$-Zr, the GB width growth at the expense of $\alpha$-Zr, which explains the terms with $G_\alpha$  in \Cref{gamma_gb}. For $\Delta\gamma_g > 0$ only, it is thermodynamically favorable for a system to split a single-phase interface into two $\alpha$-GB and $\omega$-GB interfaces. The first $l_g$-independent term $\gamma_{\alpha g}+\gamma_{\omega g}$ in \Cref{gamma_gb} appears from the condition that for $l_g=0$, the GB energy $\Gamma_{gb}=\gamma_0$, by definition. The driving force for the grain boundary broadening is given by
		\begin{equation}
			X_l = -\frac{\partial\Gamma_{gb}}{\partial l_g} = -\left(G_d - G_\alpha\right) + \frac{\Delta\gamma_g}{\delta_g} \exp\left(-\frac{l_g}{\delta_g}\right).
		\end{equation}
		The condition $X_l = 0$ results in the equilibrium GB width
		\begin{equation}\label{leq}
			l_g^e = \delta_g \ln\left(\frac{\Delta\gamma_g}{\left(G_d - G_\alpha\right)\delta_g}\right); \quad \frac{\Delta\gamma_g}{\left(G_d - G_\alpha\right)\delta_g}\geq1.
		\end{equation}
		For small increment $l_g = l_g^e + \Delta l$, we obtain in linear approximation
		\begin{equation}
			X_l = -\frac{\Delta \gamma_g}{\delta_g}\exp\left(-\frac{l^e_g}{\delta_g}\right)\frac{\Delta l}{\delta_g} = -\left(G_d - G_\alpha\right)\frac{\Delta l}{\delta_g},
		\end{equation}
		where condition $X_l \left(l_g^e\right) =0$ was taken into account. Thus, an increase in the GB width with respect to its equilibrium value causes a negative driving force for its reduction. The GB energy for the equilibrium GB width is
		\begin{align}\label{gb_eq}
			\Gamma_{gb} = &\gamma_{\alpha g} + \gamma_{\omega g} + \left(G_d - G_\alpha\right)l_g^e + \Delta\gamma_g \exp\left(-\frac{l_g^e}{\delta_g}\right)\nonumber \\
			= &\gamma_{\alpha g} + \gamma_{\omega g} + \left(G_d - G_\alpha\right)(l_g^e + \delta_g),
		\end{align}
		where, again, for the last transformation we used condition $X_l \left(l_g^e\right) =0$.  For $\Delta\gamma_g = (G_d - G_\alpha)\delta_g$ we obtain $l_g^e = 0$ from \Cref{leq}  and from \Cref{gb_eq}
		\begin{equation}\label{approx_Ggb}
			\Gamma_{gb} = \gamma_{\alpha g} + \gamma_{\omega g} + \Delta\gamma_g = \gamma_0,
		\end{equation}
		which is consistent. Note that \Cref{approx_Ggb} is a good approximation for small $l_g^e < \delta_g$ as well. Indeed,
		\begin{equation}
			\Gamma_{gb} = \gamma_{\alpha g} + \gamma_{\omega g} + (G_d - G_\alpha)\delta_g\left(\ln\left(\frac{\Delta\gamma_g}{\left(G_d-G_\alpha\right)\delta_g}\right)+1\right),
		\end{equation}
		and with $\ln\left(\frac{\Delta\gamma_g}{\left(G_d-G_\alpha\right)\delta_g}\right)\simeq\frac{\Delta\gamma_g}{\left(G_d-G_\alpha\right)\delta_g}-1$ we obtain
		\begin{equation}
			\Gamma_{gb}\simeq\gamma_{\alpha g} + \gamma_{\omega g} + \Delta\gamma_g \simeq\gamma_0,
		\end{equation}
which is consistent.
		As we will see below, this approximation just 1\% error in $\Gamma_{gb}$ even for $l_g^e = 0.5\delta_g$.

	\section{Thermodynamically equilibrium width of the $\omega$-Zr layer}

Similar consideration is valid for the barrierless appearance of the $\omega$-Zr layer of the width $l_\omega$ between the grain
boundary and ${\alpha}$-Zr. The energy of the finite-width $\omega$-Zr layer per unit area, $\Gamma_{\omega}$,   is determined by an equation similar to \Cref{gamma_gb}
		\begin{align}\label{Gamma_om}
			&\Gamma_{\omega} = \gamma_{\alpha \omega} + \gamma_{\omega g} - X_t l_{\omega} + \Delta\gamma_{\omega} exp\left(-\frac{l_{\omega}}{\delta_{\omega}}\right); \nonumber \\
			&X_t \leq 0; \quad \Delta\gamma_{\omega} = \gamma_{\alpha g}-\gamma_{\omega g}- \gamma_{\alpha\omega} > 0.
		\end{align}
Here,  $\Delta\gamma_\omega$ is the difference  in the energy between the initial $\alpha$-GB interface and the final  $\alpha-\omega$ and $\omega$-GB interfaces, which drives the PT. The exponential term describes the short-range repulsion between $\alpha$-GB and $\omega$-GB interfaces, which produces the finite width of the $\omega$ phase; $\delta_\omega$ is a characteristic interaction length. In contrast to the above consideration with the critical
$\omega$-Zr nucleus, here $\Delta\gamma_{\omega}> 0 $ (i.e., nucleation of $\omega$-Zr leads to the reduction in the total surface energy, the critical nucleus is not necessary, and barrierless equilibrium nucleation occurs; also equilibrium $\theta$ in \Cref{Het-nucl} does not exist) and $X_t \leq 0$ (otherwise, $\omega$ phase will cover the entire volume).
 The first $l_{\omega}$-independent term $\gamma_{\alpha \omega} + \gamma_{\omega g}$ in \Cref{Gamma_om} appears from the condition that for $l_{\omega}=0$, the  energy of the system $\Gamma_{\omega}=\gamma_{\alpha g}$, by definition.
The condition $X_{l \omega} = -\frac{\partial\Gamma_{\omega}}{\partial l_{\omega}} = 0$ results in the equilibrium  width of the $\omega$ phase (the counterpart of  \Cref{leq})
		\begin{equation}\label{leqom}
			l_\omega^e = \delta_\omega \ln\left(-\frac{\Delta\gamma_\omega}{X_t \delta_\omega}\right); \quad -\frac{\Delta\gamma_\omega}{X_t\delta_\omega}\geq1.
		\end{equation}
Thus, for $\Delta\gamma_{\omega}> 0 $ and $X_t<0$ but close to zero, there is always the finite-width layer  of the $\omega$ phase
at the grain boundary, even in the region of stability of the $\alpha$-Zr. By approaching thermodynamic equilibrium $X_t\rightarrow 0$,
$l_\omega^e \rightarrow \infty$, i.e., $\omega$ phase spreads over the entire grain.

	\section{Energy of the \texorpdfstring{${\alpha}-{\omega}$}{a-w} phase interface in Zr}\label{sectionS5}

		Let us estimate the energy $\gamma_{\alpha\omega}$ of the $\alpha-\omega$ phase interface in Zr using Eq. (74) from the phase field approach solution in~\citep{Levitas2003}:
		\begin{equation}
			\gamma_{\alpha\omega}=\frac{s_1}{3p_4}\delta_{\alpha\omega}={2.0 G}_{max}\delta_{\alpha\omega},
		\end{equation}
		where $G_{max}$ is the energy barrier between $\alpha$ and $\omega$ phases at the phase equilibrium, and $\delta_{\alpha\omega}$ is the width of the $\alpha-\omega$ phase interface, $s_1=16{\ G}_{max}$, and $p_4=2.667$. Note that from~\citep{Yeddu2012} $\gamma_{\alpha\omega}={1.9\ G}_{max}\delta_{\alpha\omega}$ with a slightly different definition of the phase interphase width. It follows from the first principle simulations in~\citep{Gao2016} that ${G}_{max}=0.02 eV/atom=1.9297  kJ/mol=137.941 MJ/m^3$. Taking $\delta_{\alpha\omega}=0.5\ nm$, we obtain $\gamma_{\alpha\omega}=0.138\ J/m^2$. This is a much larger value than the generic $\gamma_{\alpha\omega}=0.01 J/m^2$ used in the phase-field simulations in~\citep{Yeddu2012,Yeddu2016}; utilization of   $\gamma_{\alpha\omega}=0.138 J/m^2$ in the phase-field simulations should lead to very different and less refined microstructure.

	\section{Notes on reduction in the GB energy during the $\alpha-\omega$ PT}\label{sectionSnew}
		{Reduction in the GB energy during the $\alpha-\omega$ PT is plausible from the following point of view. The main point is that after completing direct PT, the reverse $\omega-\alpha$ PT does not occur even at complete unloading after hydrostatic loading ~\cite{Li2018a} and severe plastic deformation ~\cite{Pandey2020,Lin-Levitas-MRL-23}. This means that (a) GB suppresses nucleation of  $\alpha$-Zr due to increasing GB energy or/and (b) defects like dislocations and twins, representing stress concentrators, are not effective nucleation cites for the reverse PT. For most materials (e.g., Si~\cite{Xuan2020,yesudhas2023plastic}, steels~\cite{olson1986dislocation}, and shape memory alloys~\cite{Li2018a}) for pressure-, temperature-, and stress-induced PT, reduction in the grain size suppresses PT. The main reason is that nucleation in these materials starts at defects, like dislocations and twins. The smaller the grain size, the smaller the number of dislocations and twins in each grain, suppressing nucleation. However, for Zr, reduction in the grain size by plastic straining slightly promotes nucleation of $\omega$ phase~\cite{Kumaretal-Acta-20,pandey2023effect} and reduces the PT start pressure.
		To make it consistent with the absence of the reverse PT, this can  be explained by the  reduction in the GB energy during  $\alpha-\omega$ PT, which  promotes heterogeneous nucleation at the grain boundary. The alternative explanation used in ~\cite{Kumaretal-Acta-20} due to stress concentration at twins is less probable due to the small grain size here, while in ~\cite{Kumaretal-Acta-20}, grain size after the small plastic strain was much larger. Since the structure of the GBs is heterogeneous along the GB and different for different contacting grains, nucleation occurs at the most potent places with the smallest angle $\theta$ at corresponding pressure in \Cref{nhstresses}b,c.  This explains the very small number of nuclei in the experiment at relatively low pressure.
}

	\section{Estimation of parameters of the grain boundary }\label{sectionS6}

		A first principles study of different symmetric tilt GBs in Zr with misorientation angles in the range of $13.17^\circ$ to $27.80^\circ$ in~\cite{Plowman2022} gives the GB energy in the range of $0.51$ to $0.94\, J/m^2$ and the GB width in the range of $0.55$ to $1.55 nm$. Since the misorientation angle in nanograined material is much larger, leading to a disordering within the GB, we accept ${\Gamma}_{gb}=1\ J/m^2$ and ${l}_g^e=1\, nm$. Furthermore, we assume $\Delta\gamma_g=0.1J/m^2$ and $\delta_g=2 nm$. Then from \Cref{leq} we obtain ${G_d-G}_\alpha=30.33\times{10}^6{J}/{m^3}$, ${G_d-G}_\omega={G_d-G}_\alpha+X_t=48.40\times{10}^6{J}/{m^3}$. It follows from \Cref{gb_eq,theta_val} (i.e., without mechanics effect)  and $\gamma_{\alpha\omega}=0.138\ J/m^2$ (\Cref{sectionS5}) that $\gamma_{\omega g}=0.389\ J/m^2$ and $\gamma_{\alpha g}=0.521\ J/m^2$. Finally, from \Cref{gamma_gb}, we have $\gamma_0=1.01\ J/m^2$, which is just 1\% larger than ${\Gamma}_{gb}$. Allowing for mechanics (\Cref{new_theta} instead of \Cref{theta_val}) slightly changes $\gamma_{\omega g}=0.393\ J/m^2$ and $\gamma_{\alpha g}=0.516\ J/m^2$. Also, due to higher $X_t=46.21\times{10}^6\ J/m^3$, we obtain larger ${G_d-G}_\omega={G_d-G}_\alpha+X_t=76.54\times{10}^6{J}/{m^3}$.

\section{Thermodynamic driving force for a single grain growth}

\begin{figure}[htp]
			\centering
			\resizebox{100mm}{!}{\includegraphics[trim={90mm 70mm 90mm 60mm},clip]{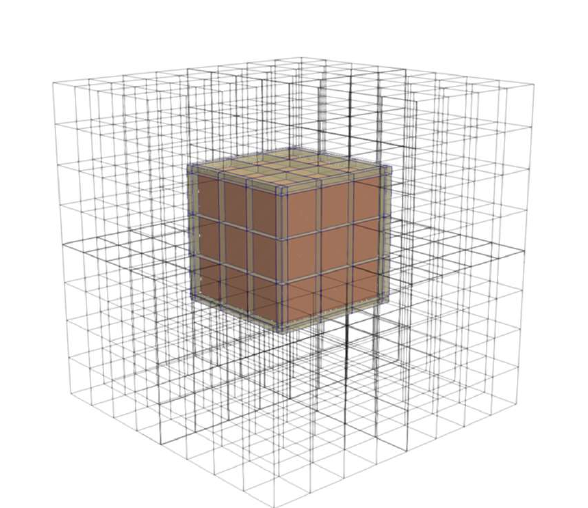}}
			\caption{\textbf{Schematic for evaluation of the thermodynamic driving force for the grain growth.} Cubic mesh represents grain structure with the size $a$ of the $\alpha$-Zr grain, which does not change in time. Equations for change in surface area and corresponding driving force are derived for the transition from the orange $\omega$-Zr cube of size $a$ to the yellow grain by absorbing parts of $\alpha$-Zr grains within an infinitesimal layer of the width $\Delta n a$ in all six directions.}
		\label{graingrowth}
		\end{figure}

	
		The dissipation rate $D$ related to the change in the total GB energy per unit GB area $A$ during grain growth with the velocity $v$ is
		\begin{equation}\label{diss}
			D = X_g v = \Gamma_{gb}\frac{\Delta A}{A \Delta t},
		\end{equation}
		where $X_g$ is the thermodynamic driving force for the GB growth per unit GB area, $\Delta A$ is the infinitesimal reduction in the GB area during infinitesimal time increment $\Delta t$. Schematics for calculating the GB area increment is shown in \Cref{graingrowth}.

		Consider the growth of an $\omega$-Zr grain with the initial size of $na$ to the final size of $(n+\Delta n)a$; $n=3$ in \Cref{graingrowth}.
		The initial surface area of 6 faces of the growing grain is $6a^2 n^2$ and the final surface area is $6a^2 (n+\Delta n)^2$, i.e., the increase is $12a^2 n\Delta n$. Reduction in the lateral area of the disappeared surrounding $\alpha$-Zr grains is $6a^2 \Delta n(n+1)^2$. Thus, the total reduction in area is
		\begin{equation}
			\Delta A = 6a^2 \Delta n[(n+1)^2-2n]=6a^2 \Delta n(2n^2+1).
		\end{equation}
		Substituting $A$ and $\Delta A$ in \Cref{diss} we obtain
		\begin{align}
			D &= X_g v = \Gamma_{gb} \frac{\Delta n \left(2n^2+1\right)}{n^2\Delta t}
				= \frac{\Gamma_{gb}}{a}\left(2+\frac{1}{n^2}\right)\frac{a\Delta n}{\Delta t} = \frac{\Gamma_{gb}}{a}\left(2+\frac{1}{n^2}\right)v; \label{eq9_1} \\
			X_g &= \frac{\Gamma_{gb}}{a}\left(2+\frac{1}{n^2}\right)\label{eq9_2}.
		\end{align}
		For initially equal grain of $\omega$-Zr and grains of $\alpha$-Zr,  $n=1$ and $X_g=\frac{3\Gamma_{gb}}{a}$, similar to the result in~\citep{Gottstein2009}. However, with the growth of the grain of $\omega$-Zr and, consequently, $n$, $X_g$ quickly reduces with $X_g=\frac{2\Gamma_{gb}}{a}$ for large $n$.
		
		Note that when the $\omega$-Zr grain of the size $a(n-{\Delta n})$ increases its size to $an$ and reaches the existing $\alpha$-Zr grain boundary, the entire area of the $\omega$-Zr GB $6a^2n^2$ disappears during infinitesimal time increment ${\Delta t}$, leading to the infinite $X_g$. However, during the next infinitesimal size increment $\Delta na$, the above procedure \Cref{eq9_1,eq9_2} are getting valid again.

\section{Combined GB and phase interface kinetics: traditional approach}\label{sectionS8}

		Thermally activated GB velocity is described by equation~\cite{Porter1992}
		\begin{equation}\label{GB_v}
			v=v_0\exp\left(-\frac{Q}{RT}\right)\left(\exp{\left(\frac{X_{gb}}{RT}\right)}-1\right).
		\end{equation}
 		where $Q$ is the activation energy, $R=8.314 JK^{-1}mol^{-1}$ is the universal molar gas constant and $v_0$ is the pre-exponential multiplier. First, let us evaluate the activation energy for the GB motion without PT. One of the components of the driving force $X_{gb}$ is given in \Cref{drivingF}. Using ${\Gamma}_{gb}=1\ J/m^2$, initial grain size $a=2\ {\times10}^{-7} m$, and $n\gg1$, we obtain
		\begin{equation}
			X_g=10 MPa=139.9 J/mol.
		\end{equation}
		One more contribution to $X_d$ maybe related to the disappearance of dislocations during grain growth~\cite{Plowman2022},
		\begin{equation}
			X_d=0.5\rho\mu_2b^2,
		\end{equation}
		where $\rho=6\times{10}^{14}/m^2$ is the dislocation density~\cite{pandey2023effect}, $\mu_2=47.83 GPa$ (\Cref{moduli-4}) is the shear modulus of $\alpha$-Zr,  $b=0.3234 nm$ is the magnitude of the Burgers vector, which results in $X_d=1.501 MPa=20.995 J/mol$. Combining with $X_t=46.21\times{10}^6 J/m^3=646.478\, {J}/{mol}$, we obtain
		\begin{equation}
			X_{gb}=X_g+X_d+X_t=807.373\, {J}/{mol}.
		\end{equation}

		To determine  $Q$ and $v_0$, we will use two conditions for the grain growth without PT, i.e., for  $X_{gb}=X_g+X_d=160.895\, {J}/{mol}$. First, as it follows from~\citep{pandey2023effect}, grain size at $T=673 K$ grows by 250 nm within 2 hours of annealing, leading to $v=0.0347 nm/s$. Second, at room temperature $T=295 K$, visible grain growth is not observed over 10 years~\citep{edalati2022nanomaterials}. Significantly overestimating, we assume that it corresponds to $v=10\,{nm}/{year}=3.215\times{10}^{-7} nm/s$. Substituting these data in \Cref{GB_v}, we obtain
		\begin{equation}
			v_0 = 19.48 \mu m/s; \quad Q = 54395.4 J/mol.
		\end{equation}
		Then including at $T=295\ K$ the additional driving force for PT, i.e., for $X_{gb}=807.373{J}/{mol}$ , we obtain $v=1.849\times{10}^{-6}\ nm/s$. It is clear that with traditional activation energy for the GB motion, no additional driving force can provide visible grain growth at room temperature. Thus, new mechanisms are required. They should be related to PT, because PT can provide the observed growth rate at room temperature.

\section{X-ray diffraction experiments}

	\subsection{Experiment I}\label{9.1}

		\subsubsection{Summary of results}

\begin{figure}[htp]
			\centering
			\resizebox{100mm}{!}{\includegraphics{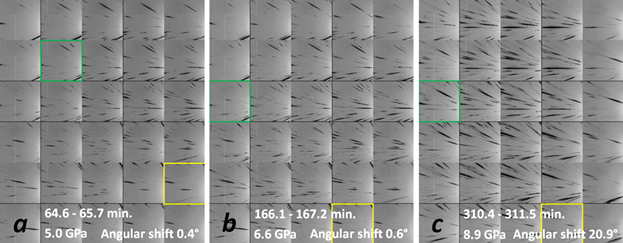}}
			\caption{\textbf{Composite frames of Laue diffraction patterns derived from the previously reported data~\cite{popov2019real} (a) and from the data collected on the same sample at higher pressures (b and c).} Time intervals when the data was collected are specified, starting from the onset of enlargement. All obtained frames are compiled as a movie (Movie S1). Yellow and green rectangles denote positions where grains 1 and 2 were located respectively. Map areas in b and c are shifted one step up and right with respect the area in a in order to follow shift of the sample due to shrink of gasket material. Intensity scaling in (a-c) is the same in Fit2d notation~\cite{hammersley1996two}.}
		\label{S1}
		\end{figure}

		As reported in the previous publication, based on these data, unusual grain enlargement of $\omega$-Zr was observed because this process caused a gradual increase of intensities of Laue reflections~\cite{popov2019real}. The most prominent enlargement of $\omega$-Zr-crystals occurred about 66 minutes after reflections from $\omega$-Zr came up. The pressure was increased from 4.1 to 5.0 GPa during this period. During further increase of pressure grain enlargement process was notably slower. Laue patterns obtained at about 167 minutes after the onset of enlargement exhibited some increase in intensities and number of reflections compared to the patterns collected at about 66 minutes after the onset (\Cref{S1}a, b, Movie S1). However, this increase is much smaller than during the first 66 minutes. This indicates that nucleation and enlargement of $\omega$-Zr crystals between 66 and 167 minutes was much slower compared to the period of time right after the onset.

\begin{figure}[htp]
			\centering
			\resizebox{100mm}{!}{\includegraphics{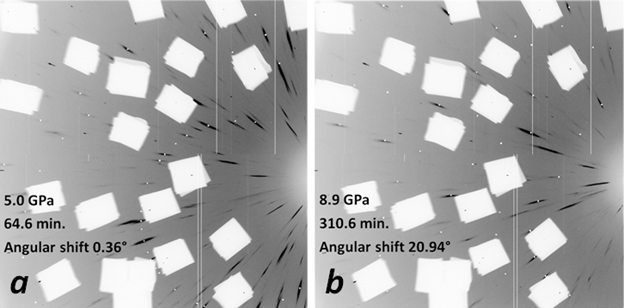}}
			\caption{\textbf{Laue diffraction patterns from an $\omega$-Zr crystal reported previously~\cite{popov2019real} (a) and collected at the highest reached pressure (b).} Times when each pattern was collected are specified, starting from the onset of enlargement. White dots denote predicted positions of reflections assuming that d-values are larger than 0.5Å. All diffraction images from this crystal are compiled as a movie (Movie S2). White areas are due to the detector mask which protected area detector from damaging by strong reflections produced by diamonds. This crystal is called grain 1. Intensity scaling in (a) and (b) is the same in Fit2d notation~\cite{hammersley1996two}.}
		\label{S2}
		\end{figure}

\begin{figure}[htp]
			\centering
			\resizebox{100mm}{!}{\includegraphics{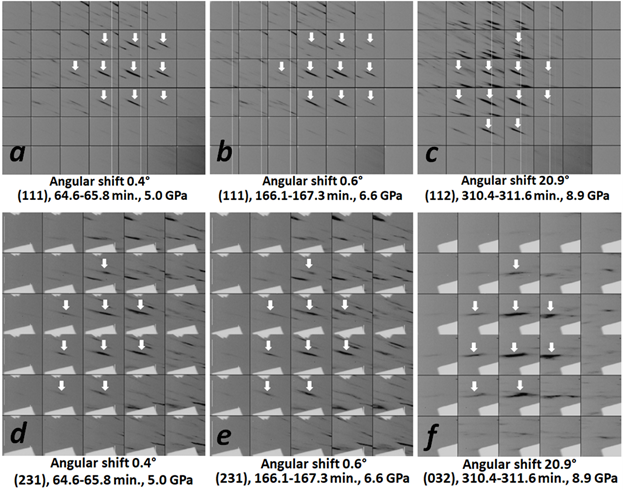}}
			\caption{\textbf{Maps of Laue reflections from $\omega$-Zr crystals called grain 1 (a-c) and grain 2 (d-f)}. Maps in a and d are derived from the previously reported data~\cite{popov2019real}. Maps in b, c, e, f have been obtained at higher pressures. White arrows denote diffraction spots produced by grains 1 and 2 in order to distinguish these spots from reflections produced by other $\omega$-Zr crystals. Time intervals when the data was collected are specified, starting from the onset of enlargement. Intensity scaling for each crystal is the same in Fit2d notation~\cite{hammersley1996two}.}
		\label{S3}
		\end{figure}

		As the sample was in contact with the gasket material via the Ruby sphere, further pressure increase caused angular shifts of the sample due to the shrink of the gasket (\Cref{Fig3}c). The highest pressure of 8.9 GPa was reached about 311 minutes after the onset of grain enlargement. Due to the angular shits, Laue patterns at the highest pressure were very different from those obtained previously (\Cref{S1}c). Therefore, comparison of sizes of $\omega$-Zr crystals at multiple pressures based directly on Laue patterns became impossible. By indexing of Laue diffraction spots, ten individual $\omega$-Zr crystals have been identified (\Cref{S2}). Enlargement processes of these crystals were evaluated based on maps of their reflections. Sample rotation caused drastic changes in diffraction intensities and, at the same time, as was demonstrated in the previous publication, intensities of Laue reflections exhibited a gradual decrease from the points of nucleation towards edges of $\omega$-Zr crystals~\cite{popov2019real}. Laue spots from edges of $\omega$-Zr crystals merge with or rise above background fluctuations if intensities of these reflections go down or go up respectively due to sample rotation. This introduces drastic changes to the area of an $\omega$-Zr crystal observed in a map of reflection without changes in crystal size. Therefore, only reflections having comparable intensities within the entire pressure range were used for mapping. Exemplarily, diffraction patterns and maps of diffraction spots from two $\omega$-Zr crystals, named grain 1, which is the same crystal as grain 1 in the previous publication~\cite{popov2019real}, and grain 2, are compiled as movies (Movies S2 and S3). Positions of these crystals within the studied area are denoted as colored rectangles in \Cref{S1}. Seven of the selected ten crystals did not exhibit notable enlargement starting from 66 minutes after the onset (\Cref{S3}).

		As it was demonstrated in the previous publication, enlargement of $\omega$-Zr crystals during the most prominent stage of this process, about an hour after the onset, was at the expense of $\alpha$-Zr rather than other $\omega$-Zr grains~\cite{popov2019real}. If the grain enlargement in Zr involved processes similar to the coarsening growth of $\omega$-Zr crystals would be at the expense of other $\omega$-Zr crystals~\cite{Doherty1997,novikov1997grain}. In this case, some of $\omega$-Zr grains would exhibit shrinkage after their enlargement. No notable shrink of 10 selected $\omega$-Zr crystals was observed in the maps of reflections within the studied pressure range. Likewise, processes similar to coarsening would cause a decrease in the overall number of reflections from $\omega$-Zr grains during compression, while intensities of remaining reflections would increase. This is not the case, as one can see from the diffraction pattern comparison in \Cref{S1}(a,b).

		At the same time, some selected negative shifts of intensities of Laue spots took place during compression. As mentioned in the previous publication, this observation may indicate minor coarsening-like phenomena, although it can also be explained by the instability of the experimental setup~\cite{popov2019real}. These shifts can also be caused by a slight movement of the sample due to shrink of the gasket, which, in turn, causes a drop in the intensities of reflections produced by edges of $\omega$-Zr crystals. For example, grain 2 exhibited a minor negative shift of diffraction intensity along with compression at 4.9GPa, which may be attributed to the slight movement of the sample causing a drop in intensities produced by the right edge and a rise of diffraction intensity produced by the left edge of the crystal (Movie S3).

		Laue diffraction spots from $\omega$-Zr crystals are quite ``streak'', indicative of a high density of crystal lattice defects. This observation is not surprising, assuming the $\omega$-Zr crystals grow from strongly disordered intergranular material. The data's spatial and time resolution limits did not allow the detection of any ``relaxation'' phenomena causing redistribution of crystal lattice defects in $\omega$-Zr crystals during their enlargement. However, such relaxation was detected on a different sample by more precise measurements (see \Cref{9.3}). 

		\subsubsection{Details of experiment and data analysis}\label{9.1.2}

		Experimental details are given in the previous publication~\cite{popov2019real}. Data reported previously were collected within the pressure range of 4.1 – 5.0 GPa in about 1 hour and 6 minutes after the onset of grain enlargement. In comparison, the overall data were collected during compression up to 8.9 GPa and in a substantially extended period of about 5 hours and 11 minutes. Here, we report the results obtained within the entire pressure range in order to investigate the post-enlargement processes. The pressure was measured right before the onset of grain enlargement after collecting reported previously and the overall data. Pressure during compression was estimated approximately through interpolation.

		Data analysis procedure included identification of $\omega$-Zr crystals by indexation and mapping of Laue spots using software routines described previously~\cite{popov2019real,popov2015high,popov2019mechanisms}. In order to evaluate changes in the overall intensities of reflections within the entire studied pressure range, composite frames of whole Laue patterns have been derived from the data collected during each translation scan (\Cref{S1}, Movie S1). Diffraction patterns in a composite frame are arranged in the same order as they were collected during the translation scan. At pressures higher than 5 GPa sample shifted due to shrink of the gasket material because the gasket was in contact with the sample via the Ruby sphere. Therefore, all map areas were shifted one step up and right in order to keep the sample centered concerning the composite frames.

		In the previous publication, four $\omega$-Zr crystals have been identified~\cite{popov2019real}. The sample did not exhibit any notable angular shifts within that pressure range. At higher pressures, Laue patterns have exhibited drastic variation across compression due to sample rotation, caused by shrink of gasket, and, as each diffraction pattern contained reflections from multiple $\omega$-Zr crystals, mapping of diffraction spots required re-identification of reflections from the same crystal by indexing at each pressure. The indexation software routine was modified accordingly to account for the angular shifts of the sample. Laue spots have been indexed at each pressure with the orientation matrix obtained at the previous pressure, assuming that the angular shift was not more significant than 30$^\circ$. This way, sets of all possible indices for each reflection were obtained. When positions and indices of two Laue reflections from the same crystal are known orientation matrix of the crystal can be calculated~\cite{chung1999automated}. Therefore, multiple combinations of positions and possible indices of pairs of reflections yielded a set of possible orientation matrices, which, in turn, have been used to try indexing all other Laue spots. When a correct combination of two reflections and their indices was used, indexing yielded all other diffraction spots from the same crystal. In this case, the number of indexed reflections was much higher than when the incorrect combination was used.

		When the first $\omega$-Zr crystal was identified, and its angular shifts at all pressures were found, there was no need to reindex reflections from other crystals at multiple pressures. Angular shifts of the first crystal were applied to other $\omega$-Zr crystals, assuming the sample was rotating as a rigid body.
		
		Orientations of all ten selected crystals match to observed Laue patterns within the entire pressure range. As examples, indexed diffraction patterns are shown in \Cref{S2} and in Movies S2 and S3. In other words, all ten selected crystals hold their relative orientations across compression. As these crystals were distributed within the sample (\Cref{Fig3}c), this indicates that the gasket material moved the sample without essential bending, which, in turn, means that the sample was not squeezed in the gasket material. Although, apart from 9 other selected crystals, grain 1 exhibits minor deviations, of less than 0.6$^\circ$, between predicted and observed orientations at pressures below 4.7 GPa, which is indicated by minor discrepancies of diffraction spots from their predicted positions in this pressure range (Movie S2). This deviation gradually reduces along with compression, indicating slight bending of the area containing grain 1 concerning the entire sample because this area was in touch with the gasket material via the Ruby sphere (\Cref{Fig3}c). Match between orientations and Laue patterns of selected crystals also indicates that the drastic changes of diffraction intensities at pressures above 5 GPa are due to changes in sample orientation caused by mechanical shifting of the sample and not due to phenomena like coarsening.
		
		It is worse to mention that the position of grain 1 slightly shifted along with compression. In contrast, the position of grain 2 stayed, meaning that the orientation of the sample shifted about a point close to grain 2. Although the angular shift of the sample reached only 20.9$^\circ$, it was not sufficient to introduce any notable changes to projections of the $\omega$-Zr crystals onto the plain of the translation scans.
		
		Shapes of Laue reflections and, therefore, types and densities of crystal lattice defects substantially varied from one $\omega$-Zr crystal to another. For example, grain 2 produced much more irregular reflections compared to grain 1. $\omega$-Zr crystals themselves were inhomogeneous in terms of distribution of crystal lattice defects, which is indicated by notable variation of shapes of diffraction spots across an $\omega$-Zr crystal. Therefore, shifts of the sample during compression may cause an increase in the broadening of reflections and, at the same time, decrease their peak and even integrated intensities if the tails of diffraction spots merge with background fluctuations. This may be a reason for minor negative shifts of diffraction intensities across compression reported in the previous publication~\cite{popov2019real}.

	\subsection{High resolution mapping of an $\omega$-Zr grain having irregular shape~\cite{popov2019mechanisms}}\label{9.2}

		\begin{figure}[htp]
			\centering
			\resizebox{100mm}{!}{\includegraphics{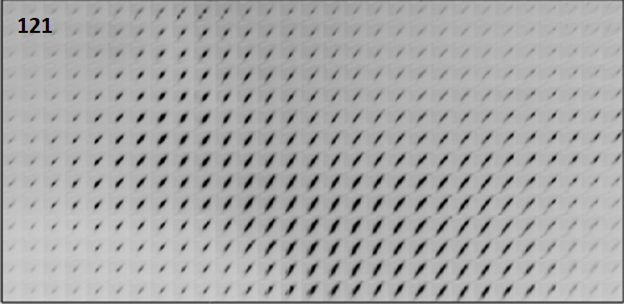}}
			\caption{\textbf{Figure 3b in~\cite{popov2019mechanisms}}. Map of a reflection from an $\omega$-Zr crystal obtained at 5.16 GPa with X-ray beam focused down to 500$\times$500 nm$^2$ (\Cref{S5}). Step size was 500 nm. }
		\label{S4}
		\end{figure}

		\begin{figure}[htp]
			\centering
			\resizebox{100mm}{!}{\includegraphics{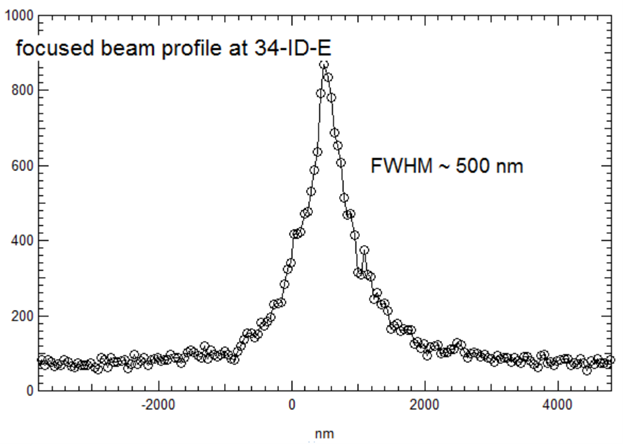}}
			\caption{\textbf{Beam profile at sample position available at beamline 34-ID-E of Advanced Photon Source.} }
		\label{S5}
		\end{figure}

	\subsection{Experiment II}\label{9.3}

		\subsubsection{Summary of the results}

		The grain enlargement process across the $\alpha\rightarrow\omega$ transition in Zr was observed with much better spatial resolution than previously, focusing on one $\omega$-Zr crystal~\cite{popov2019real} (Movie S4). The studied crystal gradually enlarged in about 2 hours and 32 minutes and stayed stable during further compression. This observation is in agreement with the data that was partially reported previously~\cite{popov2019real}. However, despite current and reported results obtained with a similar average pressure rate of about 0.8 GPa/hour, enlargement of $\omega$-Zr crystals reported previously was mainly completed in a much shorter period of about 1 hour and 6 minutes. This difference is probably caused by somewhat different distribution of residual strain, dislocations, grain sizes, and intergranular structure in the original samples of $\alpha$-Zr because scratching off samples for compression from the bulk sample was done manually without taking care about the reproducibility of their microstructure.
		
		Another possible reason why the enlargement of the $\omega$-Zr crystal took more than two times longer time than the enlargement of the crystals reported previously is that for the current data, pressure was measured directly only before and after the entire compression process, which was much longer compared to the process of enlargement itself. Therefore, the pressure rate during enlargement could differ substantially from the average rate. The pressure range of the enlargement process was roughly estimated by an approximation based on the time interval when enlargement was observed and found to be 6.3--8.4 GPa. Single crystals of Ne have been observed when the estimated pressure reached 7.2 GPa. As Ne crystallizes at 4.8 GPa, the absolute pressure was at least higher than this value~\cite{klotz2009hydrostatic}. Approximated pressures are somewhat higher than the pressure range of 4.1--5.0 GPa reported previously, measured directly before and after the enlargement process~\cite{popov2019real}. This difference may also be attributed to microstructural differences between the samples. However, currently reported pressures may be overestimated as well.
		
		It is important to stress that no shrink of the studied crystal was observed but only enlargement and stabilization. A shrink would indicate phenomena similar to coarsening when the growth of $\omega$-Zr crystals would be at the expense of other $\omega$-Zr crystals~\cite{Doherty1997,novikov1997grain}.
		
\begin{figure}[htp]
			\centering
			\resizebox{100mm}{!}{\includegraphics{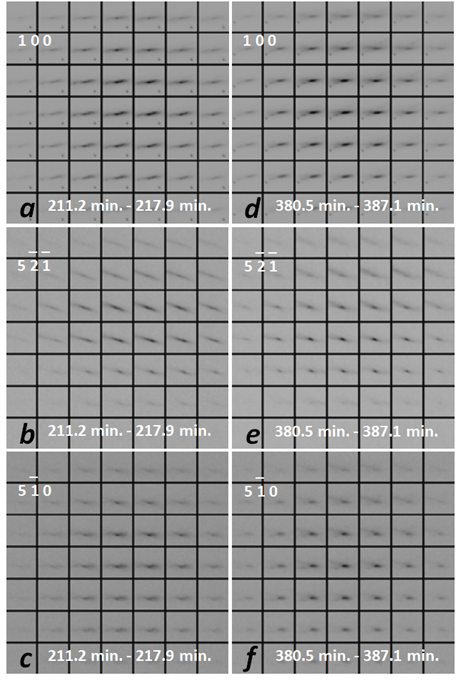}}
			\caption{\textbf{Maps of Laue reflections when the $\omega$-Zr crystal stabilized (a-c) and later (d-f)}. Intensity scaling is kept the same for each reflection in Fit2d notation~\Cref{S2}. Time intervals when each translation scan was collected are specified, starting from the onset of enlargement. The map area is outlined as blue rectangles on the slides presenting these two translation scans in movie S4.  }
		\label{S6}
		\end{figure}

		Apart from the previously reported data, due to much better spatial resolution, some release of the $\omega$-Zr crystal from defects was observed after the crystal size stabilized. Almost three hours after the enlargement process was completed, some Laue reflections became notably sharper, indicating that the density of dislocations or other defects reduced (\Cref{S6}).

		\subsubsection{Details of experiment and data analysis}

\begin{figure}[h]
		\begin{minipage}{0.45\textwidth}
			\centering
			\resizebox{70mm}{!}{\includegraphics{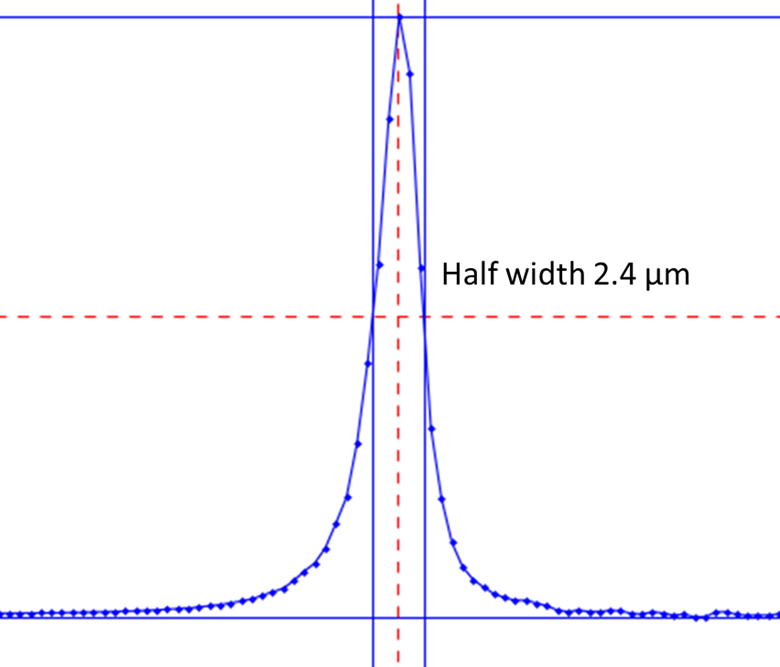}}
			\caption{\textbf{Horizontal beam profile.} }
		\label{S7}
		\end{minipage}
		\begin{minipage}{0.45\textwidth}
			\centering
			\resizebox{70mm}{!}{\includegraphics{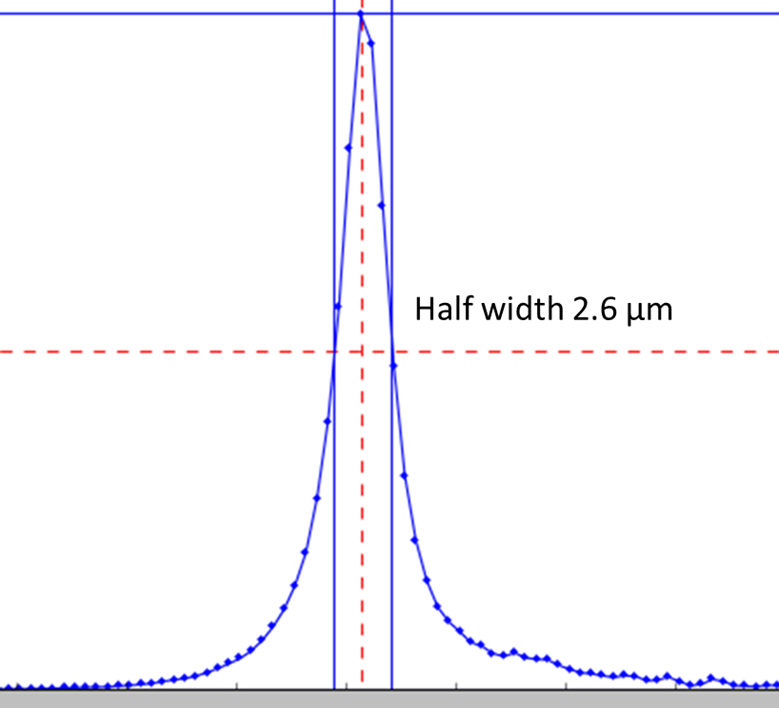}}
			\caption{\textbf{Vertical beam profile.}  }
		\label{S8}
		\end{minipage}
		\end{figure}

		The experimental setup used this time provided a much more stable X-ray beam profile and had much more stable mechanical stages than the setup used to collect previously reported data~\cite{popov2019mechanisms}. Laue diffraction measurements have been conducted using the transmission-mode geometry with the area detector tilted vertically by 30$^\circ$. X-ray beam having energy range of 10--70 keV was focused with KB-mirrors down 2.4 $\mu$m horizontally and 2.6 $\mu$m vertically at half width maximum of the beam profiles (\Cref{S7,S8}).
		
		Sample of $\alpha$-Zr for compression in a diamond anvil cell was obtained the same way as for the previously reported results by scratching off from the same bulk sample~\cite{popov2019real}. Ne was used as a pressure-transmitting medium, providing hydrostatic compression within the pressure range of interest. A series of two-dimensional translation scans, in horizontal and vertical directions, was collected on the sample across compression from 0.9 to 11.51 GPa. The pressure was increased remotely using a membrane system. The pressure was measured before and after collecting data using the off-line ruby fluorescence system in the experimental hutch~\cite{popov2019mechanisms}. The minor possible translation step of 0.5 $\mu$m, defined by the precision of mechanical stages, was implemented. Despite this step size being about five times smaller than the beam size, the interface shift between an $\omega$-Zr crystal and $\alpha$-Zr comparable to the step size still could be indicated by a drop or rise of diffraction intensities.
		
		The total scan area was 10$\times$10 $\mu$m$^2$. The sample was aligned in an X-ray beam using absorption scans with a photodiode located downstream from the sample. Taking the sample away from the setup to measure pressure would make recovering the same scan area when mounted back extremely challenging. Therefore, the pressure was not measured during compression directly, but, instead, the pressure was roughly estimated by its approximation, based on the time and assuming that the pressure rate was the same within the entire pressure range.
		
		The data analysis procedure included the identification of $\omega$-Zr crystals by indexation and mapping Laue spots using software routines described previously. Four $\omega$-Zr crystals have been identified within the scan area. Only one of the identified crystals stayed within the scan area in the entire pressure range; therefore, this crystal was selected for this study. A series of Laue patterns and maps of a reflection from this crystal are compiled in a movie (Movie S4). Three other crystals exhibited only partial overlap with the scan area. Although all four identified $\omega$-Zr crystals exhibited the same random shifts of their positions caused by the mechanical movement of the sample due to compression. Map areas in Movie S4 have also been shifted to compensate for those movements and keep the center of the crystal coincident within the entire pressure range.
		
		Two single crystals of Ne have been identified by indexing their reflections. These crystals overlapped about half of the scan area when they were observed first. Across further compression single-crystals of Ne exhibited deformation indicated by gradual change of shapes of their reflections, encircled in blue in Movie S4. These reflections became more ``streaky'' across compression, and when pressure approached the highest value, the crystals of Ne produced only very irregular ``streaks'' and not sharp diffraction spots (Movie S4).

	\subsection{Experiment III}\label{9.4}

		\subsubsection{Slow compression to the phase equilibrium pressure}

\begin{figure}[htp]
			\centering
			\resizebox{70mm}{!}{\includegraphics{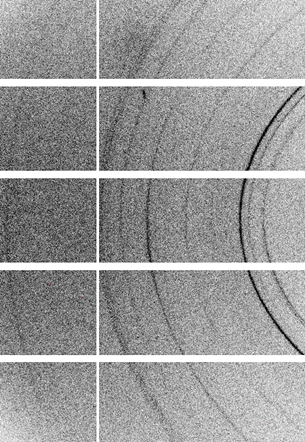}}
			\caption{\textbf{Diffraction pattern of $\alpha$-Zr at 3.24 GPa.}}
		\label{S9}
		\end{figure}

		Sample of $\alpha$-Zr for compression in a diamond anvil cell was obtained the same way as for the previously reported results by scratching off from the same bulk sample~\cite{popov2019real}. After Ne loading the diamond anvil cell, as pressure transmitting medium provided hydrostatic compression, the pressure was measured with Ruby system and was 0.34 GPa~\cite{mao1986calibration}. The pressure was further increased with the membrane system in 3 days and 6 hours, along with collecting Laue and monochromatic beam diffraction data at beamline 12.3.2 of the Advanced Light Source~\cite{kunz2009dedicated}. The transmission-mode geometry with Pilatus1M area detector, tilted vertically by 45$^\circ$, was used. X-ray beam was focused down to 1 $\mu$m. A series of two-dimensional translation scans, covering about 30\% of the sample, have been collected across compression. The studied area also included parts of a Ruby sphere and re-gasket. The step size of Laue diffraction scans was 1 $\mu$m while scans with monochromatic beam were collected with the step of 5 $\mu$. The X-ray energy range of the polychromatic beam was 6--22 keV, while the energy of the monochromatic beam was 15 keV. Geometry calibration was done on a diffraction pattern from Re gasket, obtained with a monochromatic beam, using software Dioptas~\cite{prescher2015dioptas}. Laue spots from the Ruby sphere have been identified by indexing using previously described software routines~\cite{popov2019real,popov2015high,popov2019mechanisms}. These reflections exhibited shifts across compression, indicating that the membrane system was engaged. No Laue reflections from $\omega$-Zr crystals have been observed within the entire studied pressure range while diffraction patterns collected with monochromatic beam contained only continuous rings from $\alpha$-Zr and no diffraction lines from $\omega$-Zr (\Cref{S9}). After compression, the pressure measured with a Ruby system was 3.24 GPa.

		\subsubsection{Grain enlargement at the phase equilibrium pressure}

		The sample was kept without increasing pressure for one year and ten months and studied again with X-ray diffraction at 16BMD beamline of the Advanced Photon Source~\cite{park2015new}. The X-ray beam, having the energy of 40keV, was focused down to 4$\times$3 $\mu$m$^2$ with KB-mirrors. Mar345 Image Plate was used to record diffraction patterns. The sample was kept rotating by 36$^\circ$ during the collection of an X-ray image to observe reflections from single crystals. Diffraction images have been collected on 11 points randomly distributed within the sample.
		
		Translation step scans of the same diffraction patterns have also been collected. The scan area of 30$\times$30 $\mu$m$^2$ covered most of the sample. The step size was 2 $\mu$m. X-ray energy was 39.526 keV. The beam was focused down to 5$\times$5 $\mu$m$^2$.
		
\begin{figure}[htp]
			\centering
			\resizebox{70mm}{!}{\includegraphics{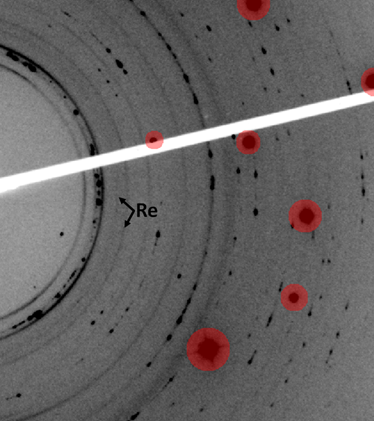}}
			\caption{\textbf{Diffraction pattern one year and ten months after the sample was compressed to 3.24 GPa.} Continues diffraction lines were produced by nano-crystalline $\alpha$-Zr except two lines from Re gasket. The 'spotty' lines were from $\omega$-Zr single-crystals. Red circles denote reflections from diamonds. }
		\label{S10}
		\end{figure}

		In all the collected images, ‘spotty’ diffraction lines from $\omega$-Zr crystals clearly dominated over continuous rings from $\alpha$-Zr (\Cref{S10}). Pressure values derived from the equations of state (EOS) of $\alpha$- and $\omega$-Zr were 3.4 GPa and 3.9 GPa respectively~\cite{anzellini2020phase}. Therefore, while the value derived from the $\alpha$-Zr data precisely matches the phase equilibrium pressure of 3.4GPa, pressure determined from the EOS of $\omega$-Zr is somewhat higher. This may be attributed to variations of the EOS due to impurities and microstructure.

		\clearpage

\section{Large grains after $\alpha\rightarrow\omega$ phase transformation in high purity Ti }
		
		\begin{figure}[htp]
			\centering
			\resizebox{150mm}{!}{\includegraphics{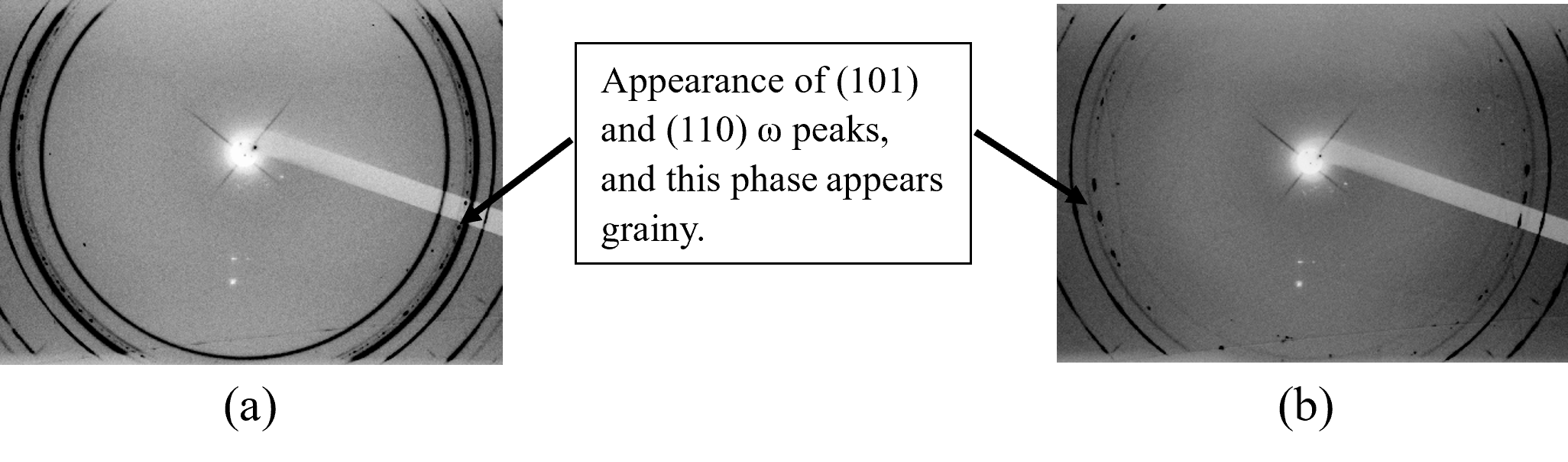}}
			\caption{\textbf{Grainy $\omega$ phase of high purity Ti that appears at the beginning of phase transformation from $\alpha$ phase under (a) nonhydrostatic compression and (b) hydrostatic compression (methanol-ethanol (4:1) pressure medium), which corresponds to the formation of large grains.}}
		\label{S12}
		\end{figure}

\section{Movie Captions}
	Movie S1. Composite frames of Laue diffraction patterns. Yellow and green rectangles denote positions where grains 1 and 2 were located respectively. Maps obtained above 5.0 GPa are shifted one step up and right with respect the maps obtained at lower pressures, in order to follow shift of the sample due to shrink of gasket material. Time intervals when the data was collected are specified, starting from the onset of $\omega$-Zr grain enlargement. 

	Movie S2. Laue diffraction patterns (left) and maps of reflections (right) from grain 1. Times when each pattern was collected are specified, starting from the onset of $\omega$-Zr grain enlargement. Yellow rectangles in the diffraction patterns outline areas around diffraction spots to build maps. Yellow rectangles in the maps denote positions where grain 1 was located. Maps obtained above 5.0 GPa are shifted one step up and right with respect the maps obtained at lower pressures in order to follow shift of the sample due to shrink of gasket material. Only reflections having comparable intensities are chosen for mapping in order to observe changes of the grain size. 

	Movie S3. Laue diffraction patterns (left) and maps of reflections (right) from grain 2. Times when each pattern was collected are specified, starting from the onset of $\omega$-Zr grain enlargement. Green rectangles in the diffraction patterns outline areas around diffraction spots to build maps. Green rectangles in the maps denote positions where grain 2 was located. Maps obtained above 5.0 GPa are shifted one step up and right with respect the maps obtained at lower pressures in order to follow shift of the sample due to shrink of gasket material. Only reflections having comparable intensities are chosen for mapping in order to observe changes of the grain size. 

	Movie S4. Maps of 100 reflection (left) and Laue patterns (right). Time intervals when each translation scan was collected are specified, starting from the onset of $\omega$-Zr grain enlargement. Yellow rectangles in diffraction patterns outline areas around diffraction spots to build maps in the right. Yellow rectangles in the maps denote positions where patterns in the right were collected. Blue rectangles outline area and denote translation scans presented in \Cref{S6}. Reflections from Ne are circled in blue. 

	For all movies: Intensity scaling of diffraction patterns is kept the same in all the slides within each movie in Fit2d notation\citep{hammersley1996two}. White circles in the diffraction patterns are due to the detector mask which protected area detector from damaging by strong reflections produced by diamonds. White dots denote predicted positions of reflections assuming that the d-values are larger than specified limits.



\end{document}